\PassOptionsToPackage{square,numbers}{natbib} 
\documentclass[10pt]{article}

\usepackage{graphicx}
\usepackage{dcolumn}
\usepackage{bm}
\usepackage{xcolor}
\usepackage[utf8]{inputenc}
\usepackage[T1]{fontenc}
\usepackage{mathptmx}
\usepackage{etoolbox}
\usepackage[normalem]{ulem}
\usepackage{amsmath}
\usepackage{appendix}

\usepackage[utf8]{inputenc}
\usepackage[T1]{fontenc}
\usepackage[english]{babel}
\usepackage{graphicx}
\usepackage{amsmath,amssymb,amsbsy,amsthm,color,natbib}
\usepackage{multicol}
\usepackage{color} 
\usepackage{natbib}
\usepackage[linktoc=all,colorlinks=true,citecolor=blue,linkcolor=blue]{hyperref}
\usepackage[paperwidth=18cm,paperheight=29.7cm,hmargin=2.5cm,vmargin=2.5cm]{geometry}
\usepackage{authblk}
\usepackage{comment}
\usepackage{cleveref}
\usepackage{subcaption}
\graphicspath{{figures/}}
\title{Energy transport and chaos in a one-dimensional disordered nonlinear stub lattice}

\author{Su Ho Cheong}
\affil{ 
Nonlinear Dynamics and Chaos Group, 
Department of Mathematics and Applied Mathematics,
University of Cape Town, Rondebosch, 7701 Cape Town, South Africa}
\affil{
Laboratoire d'Acoustique de l'Universit\'{e} du Mans (LAUM), UMR 6613, Institut d'Acoustique - Graduate School (IA-GS), CNRS, Le Mans Universit\'{e}, Av.~O.~Messiaen, F-72085 LE MANS Cedex 9, France}

\author{Arnold Ngapasare}
\affil{ 
Nonlinear Dynamics and Chaos Group, 
Department of Mathematics and Applied Mathematics,
University of Cape Town, Rondebosch, 7701 Cape Town, South Africa}

\author{Vassos Achilleos}
\author{Georgios Theocharis}
\author{Olivier Richoux}
\affil{
Laboratoire d'Acoustique de l'Universit\'{e} du Mans (LAUM), UMR 6613, Institut d'Acoustique - Graduate School (IA-GS), CNRS, Le Mans Universit\'{e}, Av.~O.~Messiaen, F-72085 LE MANS Cedex 9, France}

\author{Charalampos Skokos}
\affil{ 
Nonlinear Dynamics and Chaos Group, 
Department of Mathematics and Applied Mathematics,
University of Cape Town, Rondebosch, 7701 Cape Town, South Africa}
\affil{Max Planck Institute for the Physics of Complex Systems, N\"othnitzer Str.~38, 01187 Dresden, Germany}

\begin{document}

\maketitle

\begin{abstract}
We investigate energy propagation in a one-dimensional stub lattice in the presence of both disorder and nonlinearity. In the periodic case, the stub lattice hosts two dispersive bands separated by a flat band; however, we show that sufficiently strong disorder fills all intermediate band gaps. By mapping the two-dimensional parameter space of disorder and nonlinearity, we identify three distinct dynamical regimes (weak chaos, strong chaos, and self-trapping) through numerical simulations of initially localized wave packets. When disorder is strong enough to close the frequency gaps, the results closely resemble those obtained in the one-dimensional disordered discrete nonlinear Schr\"{o}dinger  equation and Klein-Gordon lattice model. In particular, subdiffusive spreading is observed in both the weak and strong chaos regimes, with the second moment $m_2$ of the norm distribution scaling as $m_2 \propto t^{0.33}$ and $m_2 \propto t^{0.5}$, respectively. The system's chaotic behavior follows a similar trend, with the finite-time maximum Lyapunov exponent $\Lambda$ decaying as $\Lambda \propto t^{-0.25}$ and $\Lambda \propto t^{-0.3}$.  For moderate disorder strengths, i.e., near the point of gap closing, we find that the presence of small frequency gaps does not exert any noticeable influence on the spreading behavior. Our findings extend the characterization of nonlinear disordered lattices in both weak and strong chaos regimes to other network geometries, such as the stub lattice, which serves as a representative flat-band system.
\end{abstract}

\maketitle

\section{Introduction}
\label{sec:intro}

Flatbands (FBs), correspond to dispersionless energy bands generated from the so-called compact localized states (CLS) built upon the principles of destructive interference and lattice symmetry \cite{ FlachSergej2014Dfbi, Wu}. Flatband networks (with at least one FB) have established a framework for exploring fascinating energy transport phenomena in multidimensional dynamic systems. There has been a remarkable interest in recent years on this topic, especially with tight-binding model based configurations \cite{MukherjeeSebabrata2015OoaL,Travkin,Danieli}. Furthermore, the wave-physics of FBs has lately been  investigated using mechanical lattices \cite{perchikov2017flat}, in photonic crystal waveguide networks and optical waveguide arrays \cite{ShenR.2010SDcw,PhysRevLett.114.245503, XiaShiqi2018UFLS} and in electronic systems such as superconducting wire networks and nano-engineered atomic lattices on metallic surfaces \cite{QiuWen-Xuan2016DALL,DrostRobert2017Tsie}. 

Since FBs are quite rare in nature, the pursuit of structures exhibiting FB properties has become a focus of research. Nowadays, such materials are intentionally engineered \cite{RamachandranAjith2017CFBE, RöntgenM.2018Clsa}. The sensitivity of these systems can lead to rapid fluctuations in their properties, especially when subjected to disorder, which is an inherent aspect of experimental setups involving FBs. Disorder can appear  during the fabrication of FB materials, potentially arising from inherent structural defects. The impact of various perturbations, including disorder and nonlinearity, on the diamond lattice, albeit without considering chaoticity was explored in Ref.~\cite{LeykamDaniel2013Alan}. Other interesting phenomena including what is dubbed topological inverse Anderson localization has also been studied for FB lattices \cite{InverseAnderson}. Moreover, the effect of nonlinearity was explored on FB lattices by fixing the onsite terms to zero, while examining certain forms of binary disorder nonlinearity \cite{JohanssonMagnus2015Ctfn}. However, studies which investigate the impact of disorder and chaos on the energy transport of systems exhibiting FBs are still missing in literature~\cite{Leykam24}.

In this paper, we study energy propagation in a prototypical flat-band lattice, the one-dimensional (1D) stub lattice, in order to explore the combined effects of disorder and nonlinearity on its chaotic dynamics. We focus on the evolution of initially localized excitations and monitor their spreading across a broad range of model parameters spanning different dynamical regimes. By systematically varying the disorder and nonlinearity strengths, we reveal three distinct regimes with characteristic dynamical behaviors: weak chaos, strong chaos, and self-trapping. Particular attention is given to the role of disorder near and beyond the point where the band gap closes, highlighting both the similarities with the 1D disordered discrete nonlinear Schr\"{o}dinger  equation system (DDNLS) and the disordered Klein-Gordon lattice model (DKG),  as well as the absence of any noticeable influence of small frequency gaps on the spreading dynamics.

The  paper is organized as follows. In Sect.~\ref{sec:model} we introduce the stub lattice model, and discuss the  different dynamical  regimes of the model. In Sect.~\ref{sec:Computational and Model Considerations} we present the numerical techniques we implement in our study, and the various quantities used to measure the wave packet's spreading and chaoticity. Then, in Sect.~\ref{sec:Numerical Results} we present the outcomes of our numerical investigations. In particular, we discuss some representative cases from the various dynamical regimes (Sect.~\ref{sec:rep_cases}), perform a global investigation of the system's parameter space,  emphasizing the transition between the different regimes (Sect.~\ref{sec:Exploring the Paramter Space}), as well as study the effect of the small frequency gaps on the dynamics  (Sect.~\ref{sec:freq_gap}). Finally in Sect.~\ref{sec:conclusion} we summarize and discuss our results.

\section{The stub lattice model}
\label{sec:model} 

\begin{figure}[h]
\includegraphics[scale=0.57]{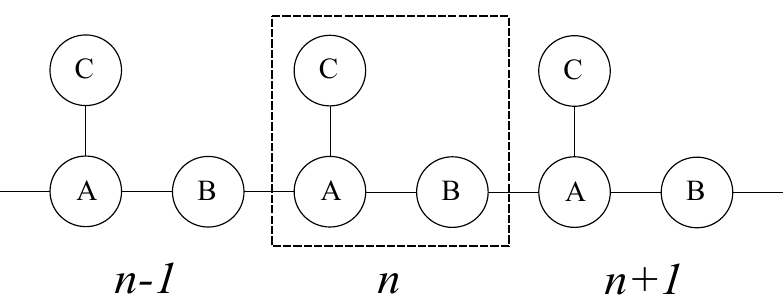}
\caption{\label{fig:schematic} Schematic representation of the stub lattice model composed of $N$ unit cells (one of which is indicated by a dashed box). Each cell contains three sites, which are denoted by circles  labeled A, B and C. The solid lines connecting the sites represent the intra- and inter-unit cell hoppings. }
\label{fig1}
\end{figure}

In our study we consider a 1D  stub lattice model with a unit cell containing three sites labeled A, B and C in the arrangement shown in Fig.~\ref{fig1}. The model exhibits only first neighbor couplings and we consider   onsite nonlinearity and onsite disorder (see e.g., Ref.~\cite{LuckJM2019Slfw}). 

The dynamics of the system can be described using a Hamiltonian formalism with 
\begin{equation}
\begin{aligned}
    H &= \sum^{N}_{n=1} \Biggl [ \epsilon_{n}^{(A)}\left|\psi_n^{(A)}\right|^{2} + \frac{\beta}{2} \left|\psi_{n}^{(A)}\right|^{4} 
    +\epsilon_{n}^{(B)}\left|\psi_n^{(B)}\right|^{2}  + \frac{\beta}{2}\left|\psi_{n}^{(B)}\right|^{4} + \epsilon_{n}^{(C)}\left|\psi_n^{(C)}\right|^{2} + \frac{\beta}{2}\left|\psi_{n}^{(C)}\right|^{4}  \\ &-  \left(\psi_n^{(C)*}\psi_n^{(A)}+\psi_n^{(C)}\psi_n^{(A)*} \right)  - \left(\psi_n^{(A)*}\psi_n^{(B)}+\psi_n^{(A)}\psi_n^{(B)*}\right)  - \left(\psi_{n+1}^{(A)*}\psi_n^{(B)}+\psi_{n+1}^{(A)}\psi_{n}^{(B)*}\right) \Biggl ],
\end{aligned}
\label{H_SL_wave}
\end{equation}
and the equations of motion are given by $\dot{\psi}_{n}^{(K)} =\partial H / \partial  (i \psi_{n}^{(K)*}) $, where $i$ is the imaginary unit, and  [$^*$] denotes complex conjugation.  
The complex variables $\psi_{n}^{(K)}$  denote the wave function amplitude at the $K$th site (with $K$ standing for A, B or C) of the $n$th cell,  and $\beta \geq 0$ represents the nonlinearity strength. The onsite disorder energies $\epsilon_{n}^{(K)} $ have uncorrelated,  randomly chosen values from a uniform distribution in the interval $[-W/2, W/2]$, with $W$ denoting the disorder strength. The stub lattice Hamiltonian [Eq.~\eqref{H_SL_wave}]  has formal similarities to the DDNLS system  with Hamiltonian $H_D=\sum^{N}_{n=1} \bigl [ \epsilon_{n} |\psi_n|^{2} + \frac{\beta}{2} |\psi_{n}|^{4} -  (\psi_{n+1}^{*}\psi_n+ \psi_{n+1}\psi_n^{*}) \bigl ]$,  whose  dynamics and wave packet spreading properties  have been studied in various works \cite{FlachS2009Usow,SkokosCh2009Dowp,LaptyevaT.V2010Tcfs,BLSKF_11,SenyangeB2018Coce}. The canonical transformation $\psi_n^{(K)} = \left( q_n^{(K)} + ip_n^{(K)} \right) /\sqrt{2}$,  $\psi_n^{(K)*} = \left( q_n^{(K)} - ip_n^{(K)} \right)/\sqrt{2}$ puts Eq.~\eqref{H_SL_wave} in the form
\begin{equation}
\begin{aligned}
H & = \sum^{N}_{n=1} \Biggl \{ \frac{\epsilon_{n}^{(A)}}{2} \left [ \left(q_n^{(A)} \right)^2+ \left( p_n^{(A)} \right)^2 \right ] 
 + \frac{\beta}{8} \left [ \left(q_n^{(A)} \right)^2+ \left( p_n^{(A)} \right)^2 \right ]^{2}  
+ \frac{\epsilon_{n}^{(B)}}{2} \left [ \left(q_n^{(B)} \right)^2+ \left( p_n^{(B)} \right)^2 \right ] \\
& + \frac{\beta}{8} \left [ \left(q_n^{(B)} \right)^2+ \left( p_n^{(B)} \right)^2 \right ]^{2}   
 + \frac{\epsilon_{n}^{(C)}}{2} \left [ \left(q_n^{(C)} \right)^2+ \left( p_n^{(C)} \right)^2 \right ] 
 + \frac{\beta}{8} \left [ \left(q_n^{(C)} \right)^2+ \left( p_n^{(C)} \right)^2 \right ]^{2} \\ 
& - \left( p_{n}^{(C)}p_{n}^{(A)} + q_{n}^{(C)}q_{n}^{(A)} \right)    
 - \left( p_{n}^{(A)}p_{n}^{(B)} + q_{n}^{(A)}q_{n}^{(B)} \right)  
 - \left( p_{n+1}^{(A)}p_{n}^{(B)} + q_{n+1}^{(A)}q_{n}^{(B)} \right) \Biggl \},
\label{H_sl_normal}
\end{aligned} 
\end{equation}
where $p_{n}^{(K)}$ and $q_{n}^{(K)}$ respectively denote the generalized momentum and position of the $K$th site in the $n$th cell. We note that the stub lattice model  conserves not only the total energy $H$ [Eq.~\eqref{H_sl_normal}] but also  the total norm defined as
\begin{equation}
    S =  \sum^{N}_{n=1} \sum_{K} \frac{1}{2} \left [ \left( q_{n}^{(K)}\right)^2 + \left( p_{n}^{(K)}\right)^2 \right ].
    \label{norm}
\end{equation}

\subsection{Energy spectrum}

\begin{figure}[h]
\includegraphics[scale=0.5]{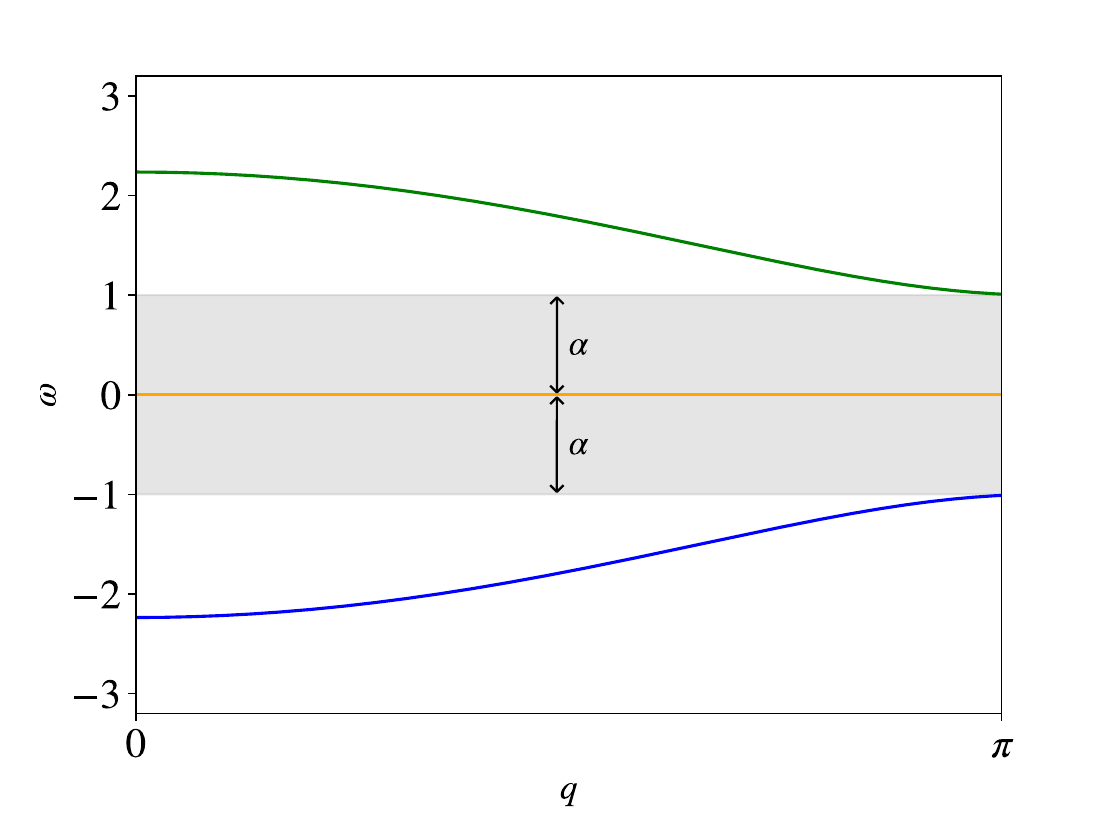}
\caption{The dispersion relation of the linear ($\beta=0$), ordered $\left ( \epsilon_{n}^{(K)} = 0 \right )$ stub lattice model \eqref{H_SL_wave}, consisting of three distinct bands: a FB (orange curve, $\omega = 0$) and two symmetric dispersive ones (green and blue curves, $\omega = \pm \sqrt{3+2\cos q}$). A gap (indicated as a gray region) of size $\alpha=1$ exists between the FB and the maximum (minimum) propagating frequency of the lower (upper) dispersive band.}
\label{fig:frequencyspectrum}
\end{figure}

Considering the solution $\psi_{n}^{(K)} = U^{(K)} \exp[-i(\omega t+qn)]$, where $q$ is the wave number,  for an infinite linear ($\beta=0$) and periodic ($\epsilon_{n}^{(K)} = 0$) lattice, we  obtain the system's dispersion relation  by expressing the frequency $\omega$ as an eigenvalue of the matrix 
\begin{equation}
\bm{E} = 
\begin{pmatrix}
0 & 1+e^{-iq} & 1\\
1+e^{iq} & 0 & 0\\
1 & 0 & 0
\end{pmatrix}.
\label{eq:E}
\end{equation} 
Thus, the spectrum consists of one FB, with $\omega = 0$ (orange curve in Fig.~\ref{fig:frequencyspectrum}) and two dispersive bands with $\omega = \sqrt{3+2 \cos q}$ (green curve in Fig.~\ref{fig:frequencyspectrum}) and $\omega = - \sqrt{3+2 \cos q}$ (blue curve in Fig.~\ref{fig:frequencyspectrum}) \cite{LeykamDaniel2017Lowd,LuckJM2019Slfw}. Between each dispersive  band and the FB there is a gap of size $\alpha=1$ (gray shaded area in Fig.~\ref{fig:frequencyspectrum}).

\begin{figure*}
\includegraphics[width=\textwidth,keepaspectratio]{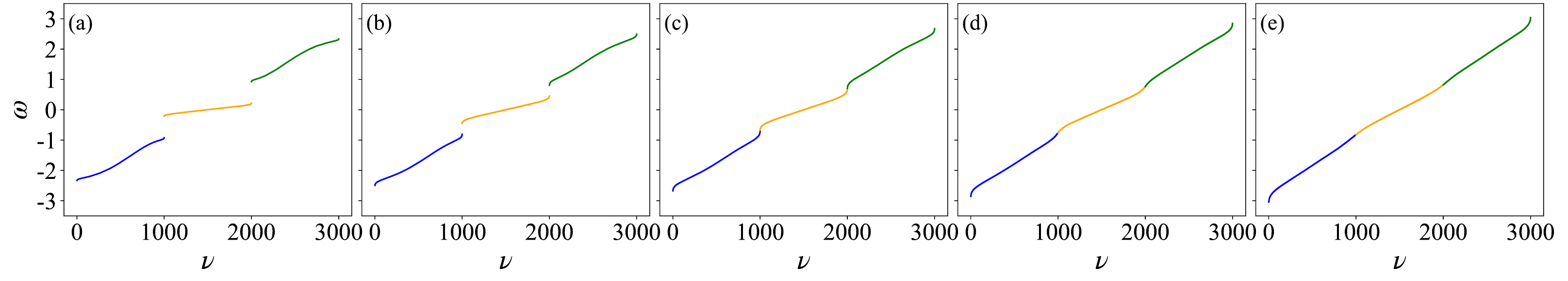} 
\caption{Averaged (over $50$ disorder realizations) frequency spectra  of the linear ($\beta=0$) disordered stub lattice model \eqref{H_SL_wave} with  $N=1000$ unit cells for different disorder strengths: (a) $W=0.5$, (b) $W=1.0$, (c) $W=1.5$, (d) $W=2.0$ and (e) $W=2.5$. In each panel the $3N=3000$ normal mode frequencies ($\omega$)  are ordered in increasing value, with  $\nu$ being the related index of this arrangement. The lowest $1000$ frequencies are colored in blue, the next $1000$ $\omega$ values  are colored in orange and the  $1000$ largest frequencies  are shown in green. We note that error bars denoting one standard deviation from the average frequency value are not present in the plots as they are too small to be visible.}
\label{fig:disorderevolution}
\end{figure*}

Before studying the dynamics, we  investigate the effect of the disorder on the frequency spectrum by considering a finite lattice of $N=1000$ unit cells ($3N$ total number of sites) with zero boundary conditions and calculating its eigenvalues. The disorder parameter $\epsilon_{n}^{(K)}$ takes random values in the interval $[-W/2, W/2]$, and we  vary the disorder strength $W$.  For each value of $W$,  we compute the average spectrum by considering 50 different random disorder realizations. The outcomes of this process for some specific values of $W$ (namely for $W=0.5$,  $W=1.0$, $W=1.5$, $W=2.0$ and $W=2.5$) are presented in Fig.~\ref{fig:disorderevolution}. We note that in this figure we arrange the frequencies in increasing order of their values and present them as a function of their index $\nu$  in this arrangement ($\nu =1,2, \ldots, 3000$). The results of Fig.~\ref{fig:disorderevolution} are colored according to the same color scheme used in Fig.~\ref{fig:frequencyspectrum} .

As shown in Fig.~\ref{fig:disorderevolution}, since any introduced disorder, however weak, destroys the symmetry of the stab lattice, the FB and the corresponding CLS are distorted. Furthermore, as $W$ increases, the band gaps decrease and eventually  disappear for $W \approx 1.58$ where  for stronger disorder strengths practically we observe a single band. Another important feature is that the total range of frequencies $\omega$ grows as $W$ increases. The maximum (minimum) permitted $\omega$ value for a specific $W$ can be  analytically found by setting all the onsite disorder energies $\epsilon_{n}^{(K)}$ to their largest (lowest) value, namely $W/2$ ($-W/2$). Thus, we find the maximum (minimum) frequency value to be $\omega_{max} = W/2 + \sqrt{5}$ ($\omega_{min} = -W/2 - \sqrt{5}$) and a total width $\Delta$ of the spectrum equal to  is
\begin{equation}
\label{eq:Delta_def}
    \Delta  =W + 2\sqrt{5}.
\end{equation}

\begin{figure*}
\includegraphics[width=\textwidth,keepaspectratio]{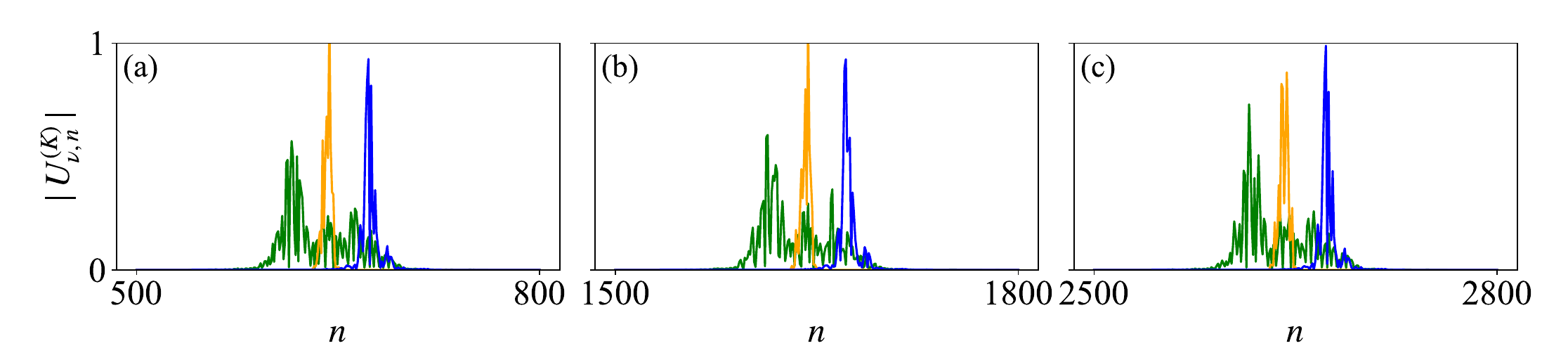} 
\caption{The profiles, i.e.~$\left| U_{\nu,n}^{(K)} \right|$ values, versus site index $n$ (with K denoting the A, B or C sites), of mode $\nu=291$ (blue), mode $\nu = 1291$ (orange), and mode $\nu = 2291$ (green) for a specific disorder realization   for the linear ($\beta = 0$) stub lattice \eqref{H_SL_wave} with $N=1000$ and $W=1$. Results are presented for  sites (a) A, (b) B, and (c) C. } 
\label{fig:mode_291}
\end{figure*}

As was already mentioned, a stub lattice of $N$ cells has $3N$ sites and, consequently,  $3N$ normal modes. Whenever a normal mode  is localized to a few lattice cells the amplitudes of the mode on the A, B and C sites of these cells will be relatively large in contrast to their values in the remaining cells. Thus, a way to present the profile of mode $\nu$, ($\nu=1,2,\ldots, 3N$) is by plotting the $3N$ absolute values of the amplitudes $\left| U_{\nu,n}^{(K)} \right|$ ($n=1,2,\ldots, N$), with $K$ denoting the A, B or C sites, for 3 ranges of $N=1000$ sites each. In this representation,  the first range corresponds to all A sites, and it is followed by the second range containing the normal mode amplitudes of all  B sites, while the results at the C sites are reported in the last range. Following this convention, we present in Fig.~\ref{fig:mode_291}, as representative cases,  the profiles of mode $\nu=291$ (blue), mode $\nu = 1291$ (orange), and mode $\nu = 2291$ (green) for a specific disorder realization  with $W=1$, in the three ranges containing results for sites A [Fig.~\ref{fig:mode_291}(a)], B [Fig.~\ref{fig:mode_291}(b)] and C [Fig.~\ref{fig:mode_291}(c)].

\subsection{Localization volume and dynamical regimes}
\label{sec:Expected Dynamical Regimes}

As shown in Fig.~\ref{fig:mode_291}, in the presence of sufficiently strong disorder, the normal modes become localized to just a few lattice sites, a phenomenon usually referred  to as Anderson localization  \cite{AndersonP.W.1958AoDi} (AL). To quantify localization, we use the so-called localization volume \cite{KF10,SenyangeB2020Ponm}, which describes the spatial extent of a mode $\nu$ by estimating the  effective distance between the exponential tails of the mode. It can be computed for each one of the sublattice sites  $K$ as $V_{\nu}^{(K)}=\sqrt{12m_{2}^{(\nu,K)}}+1$,   where $m_{2}^{(\nu,K)} = \displaystyle \sum_{n=1}^{N} \left( n-\bar{n}_{\nu}^{(K)} \right)^{2} \Big| U_{\nu,n}^{(K)} \Big|^{2}$ is the second moment of the normal mode amplitude distributions on the $K=A, \,B, \,C$ sites, with $ \bar{n}_{\nu}^{(K)} = \displaystyle \sum_{n=1}^{N} n \Big| U_{\nu,n}^{(K)} \Big|^{2} $ denoting the distribution's mean position in each sublattice. Then, for each mode, we consider the maximum of the three quantities $V_{\nu}^{(K)}$ as a good representation of its overall extent and define the mode's localization volume $V_{\nu}$ as
\begin{equation}
  V_{\nu} = \max_{K} \left( V_{\nu}^{(K)} \right) =  \max_{K} \left(\sqrt{12m_{2}^{(\nu,K)}} + 1\right).
    \label{eq:Lv}
\end{equation}
Using these definitions, we quantify the localization volume in  a wide range of disorder strengths, averaging for 
50 disorder realizations per $W$. In order to avoid boundary effects, we only keep the modes whose mean position $\bar{n}_{\nu}^{(K)}$ is located in the central one-third of the lattice and calculate the mean localization volume $\langle V^{(K)} \rangle$ leading to the corresponding average localization volume of the modes $\langle V \rangle$ being evaluated as 
\begin{equation}
  \langle V \rangle = \max_{K} \langle V^{(K)} \rangle. 
    \label{eq:average_V}
\end{equation}

\begin{figure}[t]
    \centering
    \includegraphics[scale=0.5]{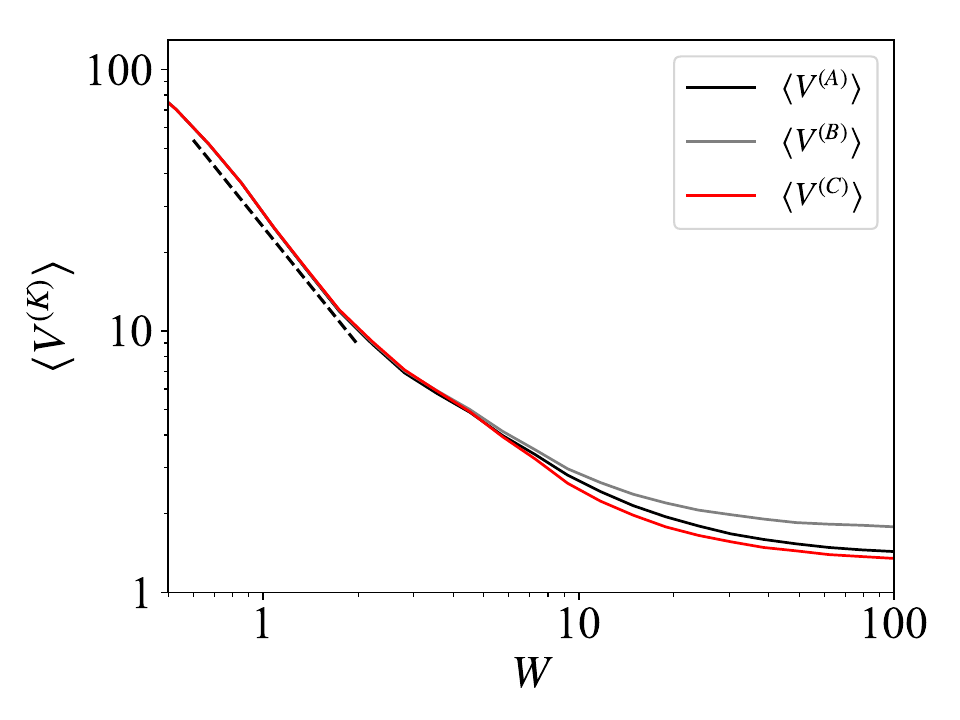}
    \caption{Averaged (over 50 disorder realizations) values  of the localization volume $\langle V^{(K)} \rangle$ on A (black curve), B (gray curve) and C sites (red curve) of normal modes with mean position at the center of the lattice, versus the disorder strength $W$. The black dashed line denotes a function proportional to $70/W^2$.
    } 
    \label{fig:volumepart}
\end{figure}

The outcomes of this process can be seen in Fig.~\ref{fig:volumepart}  where $\langle V^{(K)} \rangle$ are plotted as functions of $W$. These three quantities, and consequently $\langle V \rangle$ [Eq.~\eqref{eq:average_V}], practically coincide for $W \lesssim 5$, while $\langle V^{(B)} \rangle$ attains slightly larger values than $\langle V^{(A)} \rangle$  and $\langle V^{(C)} \rangle$ for larger $W$ values. The comparison of  Fig.~\ref{fig:volumepart} and Fig.~2 of Ref.~\cite{KF10}  show a similar behavior of $\langle V \rangle$ to the one observed for the DDNLS model along different disorder strengths. In particular,  $\langle V \rangle$ gets close to 1 for large $W$ values, while  $\langle V \rangle \approx \dfrac{70}{W^{2}}$ (dashed black line in Fig.~\ref{fig:volumepart}) for rather small disorder strengths, including the range $W \lesssim 1.58$ for which frequency band gaps are present. A similar relation (but with a  coefficient $330$ instead of $70$)  was also observed for the DDNLS system \cite{KF10}.

Based on the analogy of the stub lattice model to the DDNLS system and anticipating the influence of nonlinearity,  we identify below the various dynamical regimes of the model following arguments  similar to the ones presented in Refs.~\cite{SkokosCh2009Dowp,LaptyevaT.V2010Tcfs,FlachS.2010Sowi,LaptThesis_12,MandaBertinMany2021Ndac}. Two important scales of the linear system are the spectrum width $\Delta$ [Eq.~\eqref{eq:Delta_def}], and the average frequency spacing of normal modes $d = \Delta / \langle V \rangle$. We note that $\epsilon_{n}^{(K)} \in [-W/2, W/2]$ and that the introduction  of nonlinearity induces frequency shifts $\delta = \beta s$, with $s$ being the norm of the excited mode, leading to the interaction of the modes and the potential spreading of wave packets. 

\begin{figure}[t]
    \centering
    \includegraphics[scale=0.55]{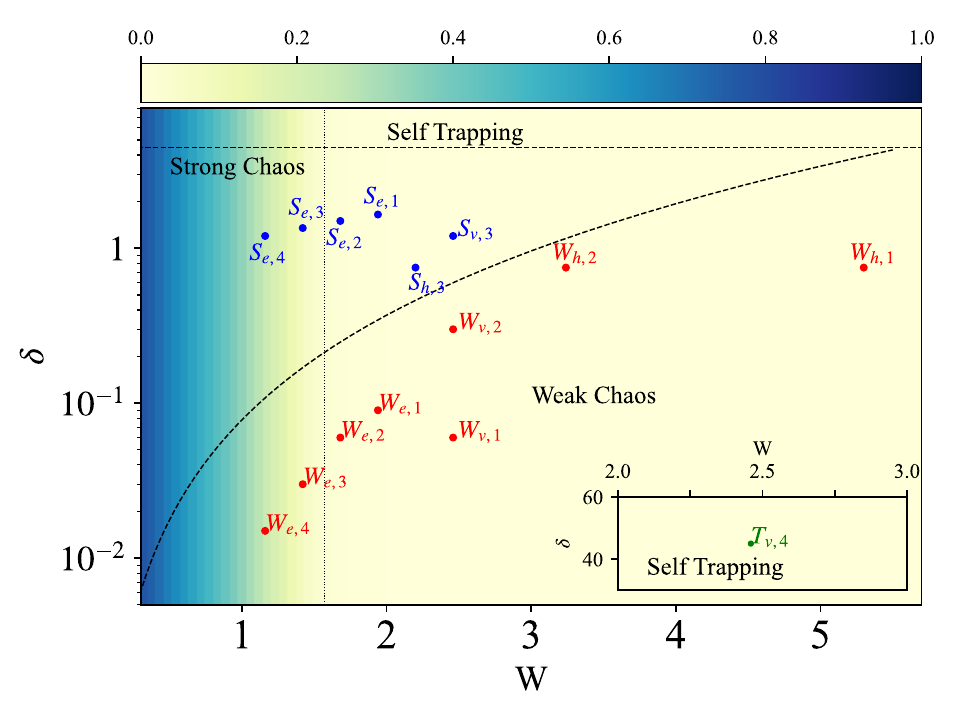}
    \caption{Parametric space of the frequency shift induced by nonlinearity ($\delta$) versus the  disorder strength ($W$), for the stub lattice model \eqref{H_sl_normal}. The color scale on the top of the panel is used for coloring each point according to the magnitude $\alpha$ of the frequency gap in the system's dispersion relation, with the vertical dotted black line at $W=1.58$ indicating the disorder strength  for which the gaps disappear (see Figs.~\ref{fig:frequencyspectrum} and \ref{fig:disorderevolution}). Three different dynamical regimes are indicated: the weak, strong and self-trapping regimes, with the boundaries between them denoted by dashed black curves corresponding to $\delta=2 \sqrt{5}$ and $\delta = \frac{W^{2}(W + 2\sqrt{5})}{70}$. The parameter values of specific cases studied in Sect.~\ref{sec:Numerical Results} are indicated by red ($W$, weak chaos), blue ($S$, strong chaos) and green  ($T$, self-trapping) points. The first subscript of each point indicates the words `vertical' ($v$), `horizontal' ($h$) or `elsewhere' ($e$), while the second numerical index is used for identifying the various cases (see text for more details). }
    \label{fig:stub_param}
\end{figure}

According to the so-called `self-trapping' theorem, \cite{KKFA08,SkokosCh2009Dowp} if the nonlinear frequency shift is large enough to exceed the spectrum width, so that some of the perturbed normal mode frequencies are tuned out of resonance with the linear normal mode spectrum, part of the wave packet will remain localized and will not spread. Actually, for a single site excitation, self-trapping will occur when $\delta \geq \Delta - 2\epsilon_n^{(K)}$. Consequently, for self-trapping to appear in the case of multiple site excitations, it is sufficient for some oscillators to be tuned out of resonance, something which can already start happening when the highest $\epsilon_n^{(K)}$ value (i.e., $W/2$) satisfy the self-trapping condition. Thus, in the parameter space $(W,\delta)$, depicted in Fig.~\ref{fig:stub_param}, the boundary for the appearance of the self-trapping regime is $\delta = \Delta - 2(W/2)=2 \sqrt{5}$ (horizontal dashed line), similar to $\delta = 2$ for the DDNLS. 

When the introduced nonlinear frequency shifts are not large  enough to induce self-trapping, initially localized wave packets can spread. If the size of the frequency shift is smaller than the average frequency spacing ($\delta < d$), modes   weakly interact, leading to a rather slow wave packet spreading in what is usually named  the `weak chaos' spreading regime, in contrast to the faster energy spreading appearing in the so-called `strong chaos' regime  for $\delta > d$ where stronger resonance overlaps can happen \cite{LaptyevaT.V2010Tcfs,BLSKF_11,SenyangeB2018Coce}. Thus, the functional expression of the boundary between the weak and strong chaos regimes in the parameter  space $(W,\delta)$ is defined by  $\delta = \frac{W^{2}(W + 2\sqrt{5})}{70}$, comparable to $\delta = \frac{W^{2}(W + 4)}{330}$ for the DDNLS system and is depicted by a  black dashed curve in Fig.~\ref{fig:stub_param}. We  note that for single site excitations we expect the appearance of only the weak chaos and self-trapping regimes as the strong chaos regime does not manifest \cite{LaptyevaT.V2010Tcfs}. We also remark that the boundaries in the system's parameter space in  Fig.~\ref{fig:stub_param} are not sharp, but rather define some transitional regions between the various dynamical regimes \cite{BLSKF_11}. 

\section{Computational techniques and model considerations}
\label{sec:Computational and Model Considerations}

In our study we follow the time evolution of initially localized excitations and analyze the characteristics of the induced wave packet propagations. We consider the normalized norm distribution
\begin{equation}
\xi_{n} = \sum_{K} \frac{ \left( q_{n}^{(K)}\right)^2 + \left( p_{n}^{(K)}\right)^2 }{2S}, \,\,\, K=1,2,3,
\label{eq:x_n}
\end{equation}
of the wave packet and compute its second moment
\begin{equation}
m_2 = \sum_{n=1}^N (n-\bar{n})\xi_{n},
\label{m2}
\end{equation}
which measures the distribution's extent. We note that in Eq.~\eqref{m2} 
\begin{equation}
\bar{n} = \sum_{n=1}^N n \xi_n,
\label{eq:mean_position}
\end{equation}
is the position of the distribution's center. We also evaluate the distribution's participation number
\begin{equation}
P = \frac{1}{\sum_{n=1}^N \xi_n^{2}}, 
\label{P}
\end{equation}
which estimates the number of the strongest excited cells and quantify the extent of the wave packet's localization.
 
We compute the finite-time maximum Lyapunov exponent (ftMLE)
\begin{equation}
\Lambda (t) =  \dfrac{1}{t}\ln \left( \dfrac{||\mathbf{w}(t)||}{||\mathbf{w}(0)||} \right)
\label{mle}
\end{equation}
to estimate the system's maximum Lyapunov exponent (MLE) $\lambda = \lim _{t\to\infty} \Lambda(t)$,  which is a commonly used measure of chaoticity \cite{S10}. In Eq.~\eqref{mle} $\mathbf{w}(0)$ and $\mathbf{w}(t)$ are, respectively, phase space deviation vectors (i.e.~vectors consisting of infinitesimal perturbations $\delta q_{n}^{(K)}$, $\delta p_{n}^{(K)}$, $n=1,2,\ldots, N$, of all generalized coordinates and momenta) from the considered orbit at $t = 0$ and $t > 0$,  while $ || \cdot ||$ denotes the usual Euclidean vector norm. In the case of regular motion, $\Lambda$ tends to zero  following the power law $\Lambda \propto t^{-1}$, while it saturates to a positive value for chaotic orbits. The time evolution of the deviation vector is governed by the so-called `variational equations' (see e.g., Refs.~\cite{SkokosCh2010Niov,S10}), which have to be solved simultaneously with the system's  equations of motion. From the evolution of deviation vectors we also compute the normalized deviation vector distribution (DVD)  
\begin{equation}
    \xi_{n}^D(t) =  
    \frac{ \displaystyle \sum_{K} \left [  \left( \delta q_{n}^{(K)}(t) \right)^{2} + \left( \delta p_{n}^{(K)}(t) \right)^{2} \right ] }
    { \displaystyle \sum_{n=1}^N \sum_{K} \left [ \left( \delta q_{n}^{(K)}(t) \right)^{2} + \left( \delta p_{n}^{(K)}(t) \right)^{2} \right ]}, 
    \label{dvd_eq}
\end{equation}
which has been used in several studies of disordered lattices to identify the spatiotemporal evolution of chaotic seeds inside the spreading wave packet \cite{SkokosCh2013Ncod,SenyangeB2018Coce,NTRSA19,MSS20}.

In our study we implement the so-called `tangent map method' \cite{SkokosCh2010Niov} to solve the stub lattice model's equations of motion and variational equations using the ABA$864$ symplectic integration scheme \cite{BlanesS.2013Nfos,DMMS19}, which proved to be very efficient for this task. Further information regarding the numerical setup and approach can be found in Appendix~\ref{APPENDIX:A1}.

In order to obtain statistically reliable results of the behavior of a quantity $R$ (like $m_2$, $P$ or $\Lambda$) we average its values over 50 different disorder realizations and smooth the obtained  results through a locally weighted difference algorithm \cite{Cleveland1988LocallyWR}, denoting the computed output as $\langle   R \rangle$. The error in the computation of any averaged, over disorder realizations, result is quantified by one standard  deviation of the averaging process. In our analysis, we usually  present the time evolution of $R$ in log-log scales and  estimate the related rate of change
\begin{equation}
    \Gamma_{R} (t) = \frac{d \langle \log_{10} R \rangle }{d \log_{10 } t}, 
\label{eq:rate}
\end{equation}
through a central finite difference, as described for example in  Ref.~\cite{LaptyevaT.V2010Tcfs}. We note that  an asymptotic  constant rate of change $ \Gamma_{R}$ implies a power law evolution of the form $R \propto t^{\Gamma_{R}}$.

\section{Numerical results}
\label{sec:Numerical Results}

\subsection{Representative cases of the various dynamical regimes}
\label{sec:rep_cases}

In order to showcase the dynamics of wave packet spreading in the different dynamical regimes of Fig.~\ref{fig:stub_param}, namely the weak chaos, strong chaos and self-trapping regime, we  numerically investigate in this section representative cases for each one of them. Based on Fig.~\ref{fig:stub_param}, we choose specific parameter setups, i.e.,~$\beta$ and $W$ values, for the stub lattice Hamiltonian  [Eq.~\eqref{H_sl_normal}] in the three distinct dynamical regimes. In our investigations, we initially excite a block of $L$ central cells of a lattice having $N$ cells, such that $L$ is practically equivalent to the average localization volume $\langle V \rangle$ [Eq.~\eqref{eq:average_V}] of the system's normal modes. The considered initial condition is defined by putting all the generalized positions of the excited part of the lattice to $q_l^{(K)} = 0$, while the conjugate momenta are randomly set to $p_l^{(K)} = \pm\sqrt{2}$. In this way the norm of each excited cell is $s = 3$. Furthermore, for all cells outside the initially excited part of the lattice, we consider $p_l^{(K)} = q_l^{(K)} = 0$. For each studied case we compute the time evolution of $\langle  \log_{10} m_2 \rangle$, $\langle \log_{10}  P \rangle$ and $\langle \log_{10}  \Lambda \rangle$, by considering 50 disorder realizations.

\subsubsection{Weak chaos spreading regime}
\label{sec:weak_rep_cases}

\begin{figure}[t]
    \centering
    \includegraphics[scale=0.5]{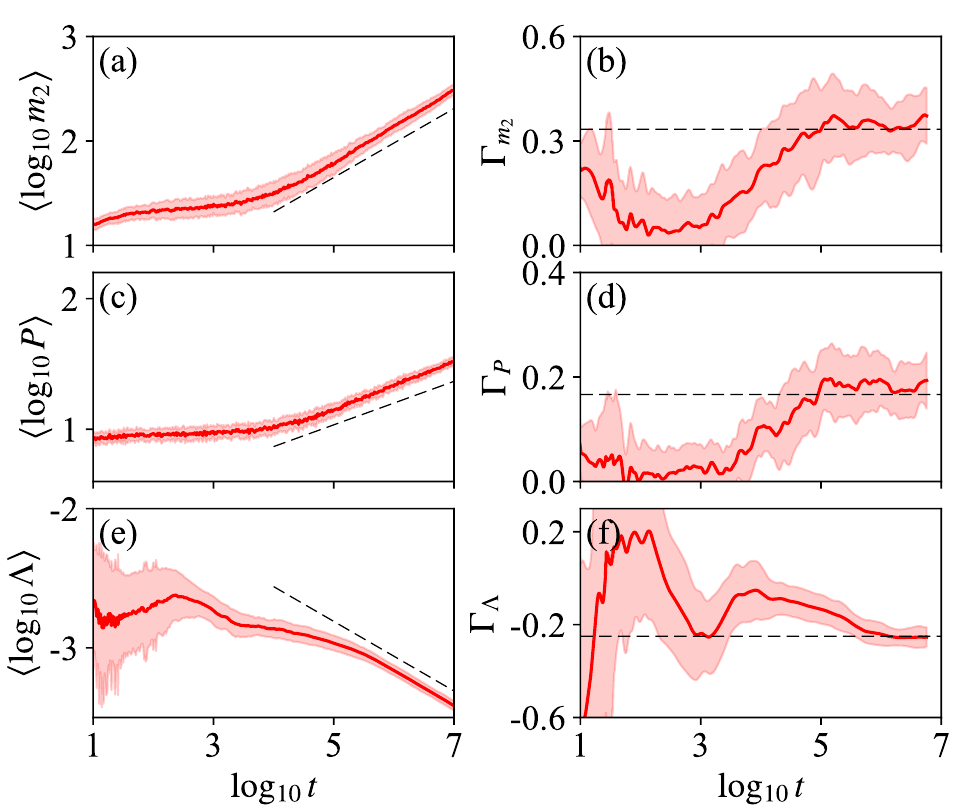}
    \caption{Averaged (and smoothed) results over 50 disorder realizations of the time evolution of the wave packet's (a) second moment $m_2(t)$ \eqref{m2}, (c) participation number $P(t)$ \eqref{P}, and (e) ftMLE $\Lambda(t)$ \eqref{mle}, along with the related numerically computed [Eq.~\eqref{eq:rate}] derivatives (b) $\Gamma_{m_2}(t)$, (d)  $\Gamma_P(t)$, and (f) $\Gamma_{\Lambda}(t)$ for the weak chaos case $W_{v,1}$ of \eqref{H_sl_normal}. The straight dashed lines correspond to values [(a) and (b)] $\Gamma_{m_2}=0.33$, [(c) and (d)] $\Gamma_{P}=0.167$, and [(e) and (f)] $\Gamma_{\Lambda}=-0.25$. In each panel the shaded area denotes one standard deviation error. }
    \label{fig:wk20}
\end{figure}

We start our investigation by considering the case
\begin{itemize}
    \item $W_{v,1}$:  $\beta = 0.02$, $W=2.46$, $L = 12$, $N=1001$,
\end{itemize}
as a representative example of the weak chaos regime. From the results of Figs.~\ref{fig:wk20}(a) and \ref{fig:wk20}(c) where respectively the time evolution of $\langle \log_{10} m_2 \rangle(t)$ and $\langle \log_{10} P \rangle(t)$  are  shown, we see that the initially localized wave packet eventually spreads, as both the second moment $m_2$ [Eq.~\eqref{m2}], and the participation number $P$ [Eq.~\eqref{P}] grow. This spreading becomes pronounced after $t \approx 10^4$, with both quantities exhibiting a power law increase. The numerically computed derivatives [see Eq.~\eqref{eq:rate}] $\Gamma_{m_2}(t)$ and $\Gamma_P(t)$ of the curves depicted in Figs.~\ref{fig:wk20}(a) and \ref{fig:wk20}(c),  are respectively given in Figs.~\ref{fig:wk20}(b) and \ref{fig:wk20}(d). Both  derivatives tend to an almost constant value, namely $\Gamma_{m_2} \approx 0.33$ and $\Gamma_P \approx 0.167$. These two values are represented by horizontal dashed lines in Figs.~\ref{fig:wk20}(b) and \ref{fig:wk20}(d), and respectively  correspond to the straight dashed lines depicted in Figs.~\ref{fig:wk20}(a) and \ref{fig:wk20}(c). It is worth noting that the numerical values of these power law exponents ($\Gamma_{m_2} = 0.33$ and $\Gamma_P = 0.167$) are very close to the exponents describing the wave packet's spreading in the weak chaos regime of both the DKG and DDNLS systems,  for which $m_2 \propto t^{1/3}$ and $P \propto t^{1/6}$ \, \cite{FlachS2009Usow,SkokosCh2009Dowp,LaptyevaT.V2010Tcfs,BLSKF_11,SkokosCh2013Ncod,SenyangeB2018Coce}.

In Fig.~\ref{fig:wk20}(e) the time evolution of the averaged  and smoothed  $\langle \log_{10} \Lambda \rangle(t)$ values is shown, while in Fig.~\ref{fig:wk20}(f) the numerically computed derivative of this curve [through the implementation of Eq.~\eqref{eq:rate}] is given. These results clearly suggest that in the weak chaos case the evolution of the ftMLE $\Lambda$ [Eq.~\eqref{mle}] is well described by  the power law $\Lambda \propto t^{-0.25}$. We note that the exponent $\Gamma_{\Lambda}=-0.25$  has also been observed for the DKG and the DDNLS systems \cite{SkokosCh2013Ncod,SenyangeB2018Coce}. The decrease of the values of $\Lambda$  in time denotes that the strength of chaos diminishes as time grows. As the wave  packet spreads, its total norm [Eq.~\eqref{norm}], which is a conserved quantity, is shared among a growing number of cells. Consequently, the strength of the nonlinear terms in Eq.~\eqref{H_sl_normal} diminishes, resulting in a decrease of the overall chaoticity  of the system. 

\begin{figure}[t]
    \centering
    \includegraphics[width=\linewidth]{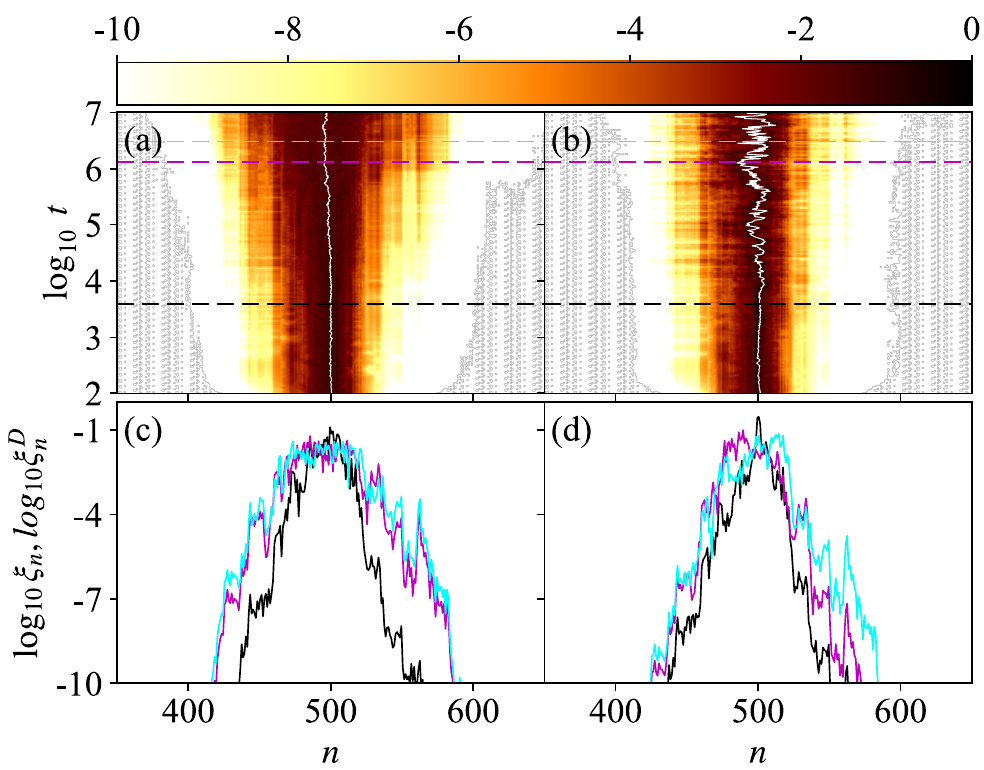}
    \caption{The dynamical behavior of a representative initial condition of the $W_{v,1}$ weak chaos case for one disorder realization. Spatiotemporal evolution of (a) the norm density $\xi_n$ \eqref{eq:x_n} and (b) the related DVD  $\xi_n^D$ \eqref{dvd_eq}. The color scale at the top of the figure is used for coloring lattice cells according to their (a) $\log_{10} \xi_n$ and (b)  $ \log_{10} \xi_n^D$. In both panels a white curve denotes the distribution’s mean position. Distributions of  $\xi_n$ and $\xi_n^D$ at times $\log_{10}t=3.64$ (black curves), $\log_{10}t=6.09$ (magenta curves) and $\log_{10}t=6.48$ (cyan curves) are respectively shown in (c) and (d). These times are also denoted by similarly colored horizontal dashed lines in (a) and (b).}
    \label{fig:wk20_map}
\end{figure}

In Fig.~\ref{fig:wk20_map}(a) [Fig.~\ref{fig:wk20_map}(b)] we see the spatiotemporal  evolution of the norm density $\xi_n$ [Eq.~\eqref{eq:x_n}] [DVD  $\xi_n^D$ [Eq.~\eqref{dvd_eq}]] of a wave packet excitation  belonging to the $W_{v,1}$ weak chaos case, for an individual disorder realization. In Fig.~\ref{fig:wk20_map}(c) [Fig.~\ref{fig:wk20_map}(d)] snapshots of this distribution taken at the instances denoted by horizontal dashed lines in Fig.~\ref{fig:wk20_map}(a) [Fig.~\ref{fig:wk20_map}(b)] are shown.  From the results of Figs.~\ref{fig:wk20_map}(a) and \ref{fig:wk20_map}(c), we see that the normalized norm distribution expands continuously to larger regions of the lattice, encompassing at the final integration time $t=10^7$ approximately 75 lattice cells to both sides of the location of the initial excitation. This spreading is done more or less symmetrically around the position of the initial excitation at the middle of the lattice, something which is reflected in the evolution of the distribution's mean position $\bar{n}$ [Eq.~\eqref{eq:mean_position}], denoted by the white  curve in Fig.~\ref{fig:wk20_map}(a). This curve is rather smooth, always remaining close to the lattice's center. On the other hand, the DVD, which always stays  inside the excited part of the lattice [Figs.~\ref{fig:wk20_map}(b) and \ref{fig:wk20_map}(d)], has a more localized shape, which does not change drastically over time. From the evolution of the DVD's mean position [white  curve in Fig.~\ref{fig:wk20_map}(b)],  we see that after an initial time interval during which it remains practically located at the  region where the initial excitation occurred, the DVD starts moving around, exhibiting fluctuations in its mean position, the amplitudes of which grow in time. Similar behaviors of the DVDs have been observed in Refs.~\cite{SkokosCh2013Ncod,SenyangeB2018Coce} for the DKG and the DDNLS systems, where the importance of the DVD oscillations as a mechanism leading to the homogenization of chaos inside the wave packet was stressed.

\subsubsection{Strong chaos spreading regime}
\label{sec:strong_rep_cases}

\begin{figure}[t]
    \centering
  \includegraphics[scale=0.5]{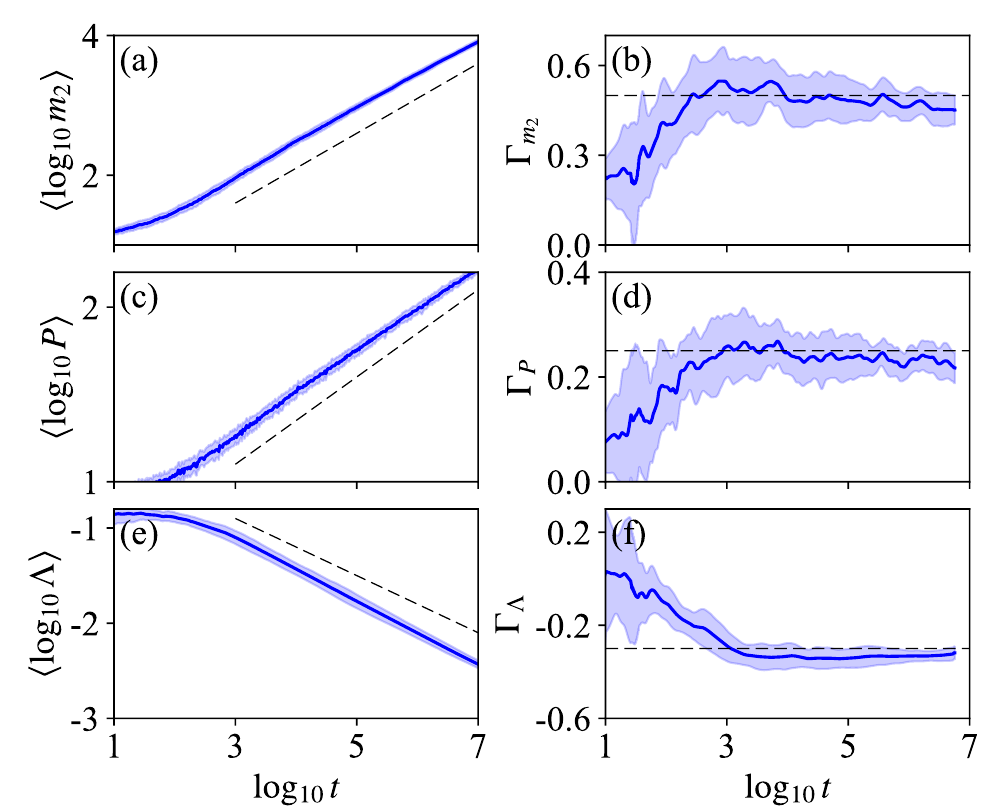}
    \caption{Similar to Fig.~\ref{fig:wk20} but for the strong chaos case $S_{v,3}$. The straight dashed lines correspond to values [(a) and (b)] $\Gamma_{m_2} = 0.5$, [(c) and (d)] $\Gamma_{P} = 0.25$, and [(e) and (f)] $\Gamma_{\Lambda} = -0.3$.}
    \label{fig:s13}
\end{figure}

\begin{figure}[t]
    \centering
    \includegraphics[width=\linewidth]{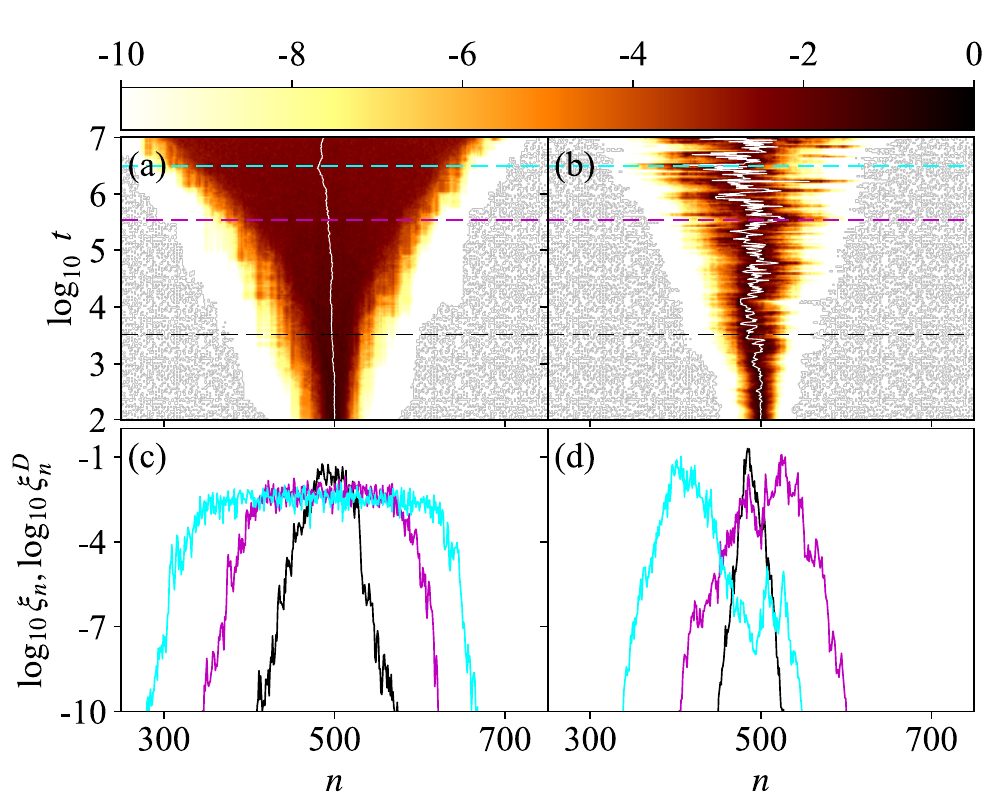}
    \caption{Similar to Fig.~\ref{fig:wk20_map} but for the strong chaos case $S_{v,3}$. The distributions in (c) and (d) are taken at  times $\log_{10}t=3.54$ (black curves), $\log_{10}t=5.53$ (magenta curves) and $\log_{10}t=6.48$ (cyan curves).}
    \label{fig:s13_map}
\end{figure}

We now turn our attention to the strong chaos spreading regime, by considering the  case
\begin{itemize}
  \item $S_{v,3}$:  $\beta = 0.4$, $W=2.46$, $L = 12$, $N=1001$.
\end{itemize}
As one can observe from the results of  Fig.~\ref{fig:s13}, in this case the wave packet spreads faster than in the weak chaos case [Fig.~\ref{fig:wk20}], as both the second moment [Figs.~\ref{fig:s13}(a) and \ref{fig:s13}(b)],  and the participation number [Figs.~\ref{fig:s13}(c) and \ref{fig:s13}(d)] show  a steeper increase. These increases are quite well approximated by  the power laws $m_2 \propto t^{0.5}$ and $P \propto t^{0.25}$, which are  denoted by straight  dashed lines in respectively Figs.~\ref{fig:s13}(a) and \ref{fig:s13}(c).  It is worth noting that these power law exponents [$\Gamma_{m_2}=0.5$ and $\Gamma_{P}=0.25$ in respectively Figs.~\ref{fig:s13}(b) and \ref{fig:s13}(d)]  also characterize the wave packet spreading in the strong chaos  regime of the DKG and DDNLS models \cite{FlachS2009Usow,SkokosCh2009Dowp,LaptyevaT.V2010Tcfs,BLSKF_11,SenyangeB2018Coce}. The faster spreading of the wave packet in the strong chaos regime, with respect to the weak chaos one, is also clearly depicted in Figs.~\ref{fig:s13_map}(a) and \ref{fig:s13_map}(c) where the evolution of a particular initial excitation for one  disorder realization belonging in the $S_{v,3}$ case is shown. More specifically, we observe again a, more or less, symmetric propagation of the wave packet on both sides of the location of the initial excitation, which is also reflected on the fact that the wave packet's mean position [white curve in Fig.~\ref{fig:s13_map}(a)] remains very close to the center of the lattice. Actually, the wave packet extends to approximately 200 lattice cells on both sides of  the initial excitation's location at $t=10^7$.

As was explained in Refs.~\cite{LaptyevaT.V2010Tcfs,BLSKF_11},
the strong chaos spreading behavior is  a transient one followed by a subsequent slowing down  of the wave packet's spreading, which asymptotically  tends to the weak chaos behavior. Signs of this transient nature of the dynamics can be seen in Figs.~\ref{fig:s13}(b) and \ref{fig:s13}(d) where  the exponents $\Gamma_{m_2}(t)$ and $\Gamma_{P}(t)$ respectively attain values close to $0.5$ and $0.25$ around $t = 10^3$, but afterwards show some slight decline to lower values, although our numerical simulations are not long enough to clearly show the  transition to the weak chaos regime  characterized  by $\Gamma_{m_2}=0.33$ [Fig.~\ref{fig:wk20}(b)] and $\Gamma_{P}=0.167$ [Fig.~\ref{fig:wk20}(d)].

Similarly to what is observed in the weak chaos case [Figs.~\ref{fig:wk20}(e) and \ref{fig:wk20}(f)] the ftMLE  exhibits a power law decay [Figs.~\ref{fig:s13}(e) and \ref{fig:s13}(f)] but with a different exponent value, namely $\Gamma_{\Lambda} \approx -0.3$. This exponent  was also found in Ref.~\cite{SenyangeB2018Coce} where  the chaotic behavior of the DKG and the DDNLS models in the strong chaos regime was studied. The related DVD [Figs.~\ref{fig:s13_map}(b) and \ref{fig:s13_map}(d)] is, as in the weak chaos case [Figs.~\ref{fig:wk20_map}(b) and \ref{fig:wk20_map}(d)],  more concentrated  than the wave packet, maintaining a rather pointy shape, and always remaining  inside the excited part of the lattice. The evolution of the DVD's mean position [white curve in Fig.~\ref{fig:s13_map}(b)],  as well as the three DVD profiles depicted in Fig.~\ref{fig:s13_map}(d), clearly show the meandering of the DVD inside the  wave packet. This behavior of the DVD  supports the homogenization of chaos inside the spreading wave packet, as was also reported in Refs.~\cite{SkokosCh2013Ncod,SenyangeB2018Coce} for the DKG and the DDNLS systems.

\subsubsection{Self-trapping regime}
\label{sec:self_rep_cases}

\begin{figure}[t]
    \centering
    \includegraphics[scale=0.5]{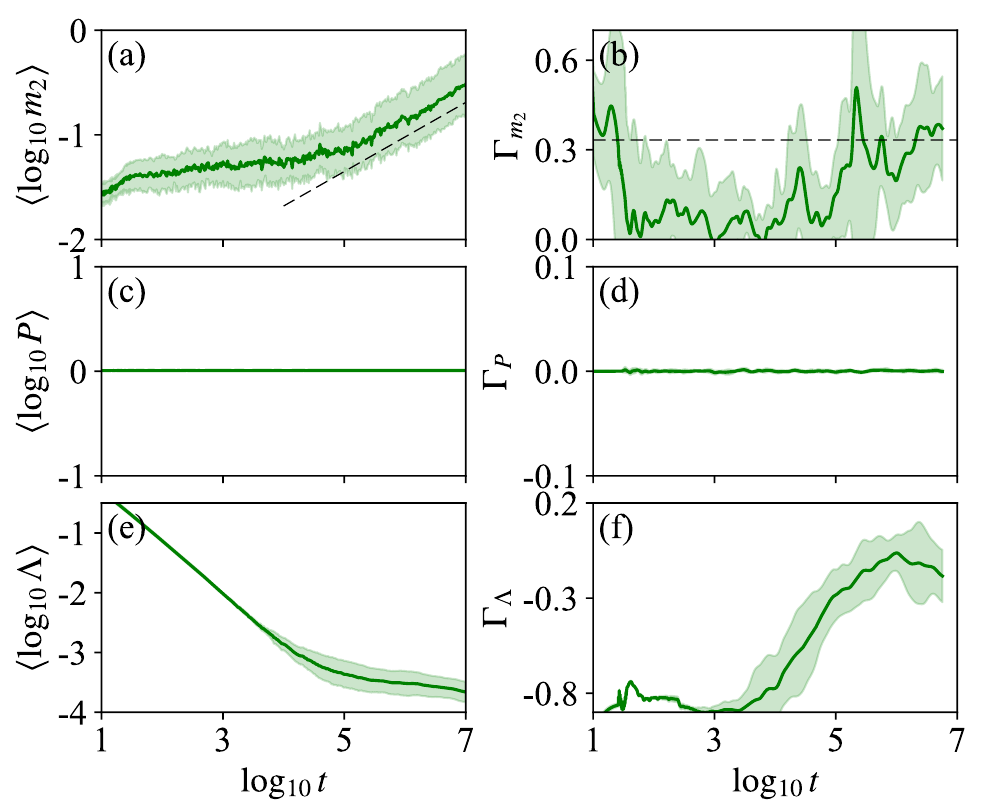}
    \caption{Similar to Fig.~\ref{fig:wk20} but for the self-trapping case $T_{v,4}$. In (a) and (b) the straight dashed lines correspond to the value $\Gamma_{m_2} = 0.33$.}
    \label{fig:sftr1}
\end{figure}

\begin{figure}[t]
    \centering
    \includegraphics[width=\linewidth]{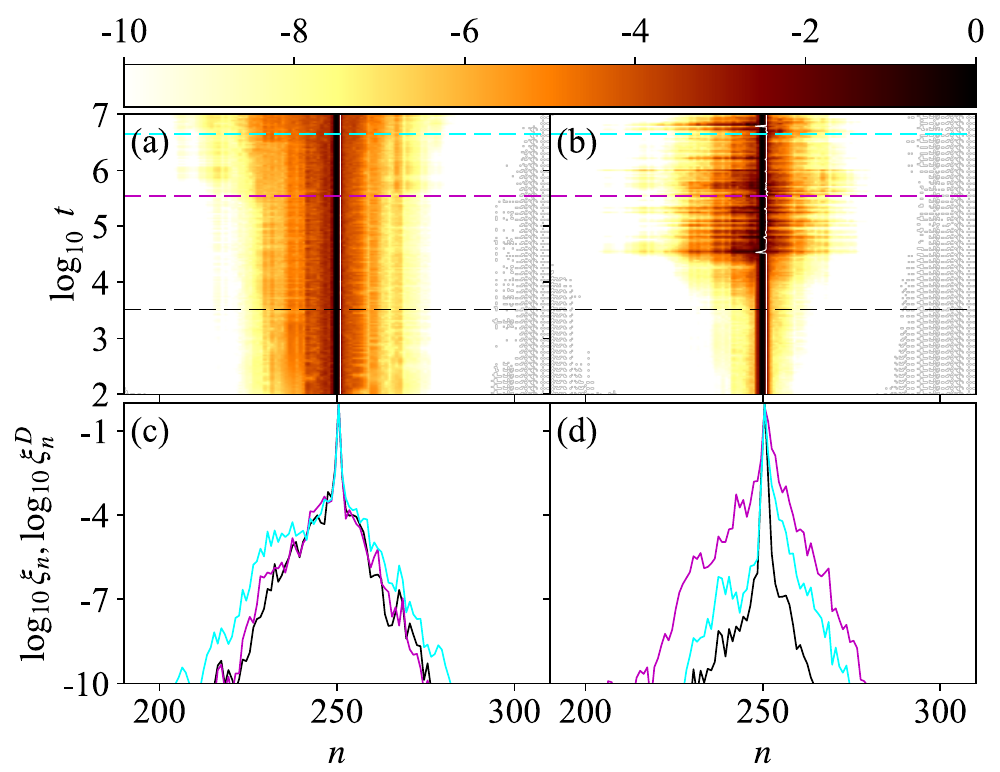}
    \caption{Similar to Fig.~\ref{fig:wk20_map} but for  the self-trapping case $T_{v,4}$. The distributions in (c) and (d) are taken at  times $\log_{10}t=3.54$ (black curves), $\log_{10}t=5.53$ (magenta curves) and $\log_{10}t=6.48$ (cyan curves).}
    \label{fig:sftr1_map}
\end{figure}

In order to showcase the dynamical behavior of the system in the self-trapping regime,  we consider the case
\begin{itemize}
  \item $T_{v,4}$:  $\beta = 15.0$, $W=2.46$, $L = 1$, $N=501$. 
\end{itemize}
As was explained in Sect.~\ref{sec:Expected Dynamical Regimes}, in the self-trapping regime the largest part of the wave packet remains localized, and consequently the participation number will stay practically constant, although some low intensity tails could propagate leading to the increase of the second moment value. This behavior, which was theoretically predicted in Refs.~\cite{KKFA08,SkokosCh2009Dowp} and numerically observed for the DKG and the DDNLS models in Refs.~\cite{FlachS2009Usow,SkokosCh2009Dowp,LaptyevaT.V2010Tcfs,BLSKF_11},  also appears in the $T_{v,4}$ case as the results of Figs.~\ref{fig:sftr1} and \ref{fig:sftr1_map} clearly show. In particular, the participation number for this case remains practically constant throughout the wave packet's evolution [Figs.~\ref{fig:sftr1}(c) and \ref{fig:sftr1}(d)] retaining its initial value, which in this case is $P=1$ as $L=1$, while the second moment starts growing after $t \approx 10^4$ [Fig.~\ref{fig:sftr1}(a)] following a power law with exponent $\Gamma_{m_2} \approx 0.33$ [Fig.~\ref{fig:sftr1}(b)].  The behavior of the $m_2$ and $P$ can be better understood by checking the time evolution of the distribution of the  norm density $\xi_n$ [Eq.~\eqref{eq:x_n}]  for one representative  self-trapping case presented in Figs.~\ref{fig:sftr1_map}(a) and \ref{fig:sftr1_map}(c). In these figures, the wave packet stays practically localized, as a high peak remains at the center of the lattice [Fig.~\ref{fig:sftr1_map}(c)], although  some very low intensity tails slightly propagate to both sides of the initial excitation. 

Despite the mainly localized behavior of the wave packet, the overall dynamics is weakly chaotic as the evolution of the  ftMLE in Fig.~\ref{fig:sftr1}(e) shows. After about $t\approx 10^4$ $\langle \log_{10} \Lambda \rangle(t)$  changes its rate of decrease, reaching very small values,  smaller than the ones observed in the weak [Fig.~\ref{fig:wk20}(e)] and strong chaos [Fig.~\ref{fig:s13}(e)] cases. Nevertheless, the  exponent $\Gamma_{\Lambda}(t)$ [Fig.~\ref{fig:sftr1}(f)], although it does not saturate to a well defined value, does not  show a tendency to approach  $\Gamma_{\Lambda}=-1$,  which is a characteristic of regular behavior \cite{S10}. The related DVD [Figs.~\ref{fig:sftr1_map}(b) and \ref{fig:sftr1_map}(d)] remains very localized at the location of the initial excitation, showing an oscillatory slight expansion of its tails [note that in Fig.~\ref{fig:sftr1_map}(d) the DVD for $\log_{10}t=5.53$ (magenta curve) is more extended than the one appearing at the later time $\log_{10}t=6.48$ (cyan curve)].

\subsection{Global investigation of the parameter space}
\label{sec:Exploring the Paramter Space}

As we mentioned at the end of Sect.~\ref{sec:Expected Dynamical Regimes}, the boundaries between the different dynamical regimes in the parameter space $(W, \delta)$ (Fig.~\ref{fig:stub_param}) of the stub lattice system  are not sharp, but rather indicate transition regions. In this section, we further investigate  the dynamical behavior of the weak and  strong chaos spreading regimes of the Hamiltonian in Eq.~\eqref{H_sl_normal}, by  studying in more detail the changes happening at the transition zone between these two regimes in the system's parameter space. In particular, we consider cases for which the disorder strength $W$ or the nonlinearity parameter $\beta$ (and consequently the frequency shift $\delta=\beta s$) is kept constant, while the other parameter is varied in such a way that we move from the weak chaos regime to the strong chaos one.

\begin{figure}[t]
    \centering
    \includegraphics[scale=0.5]{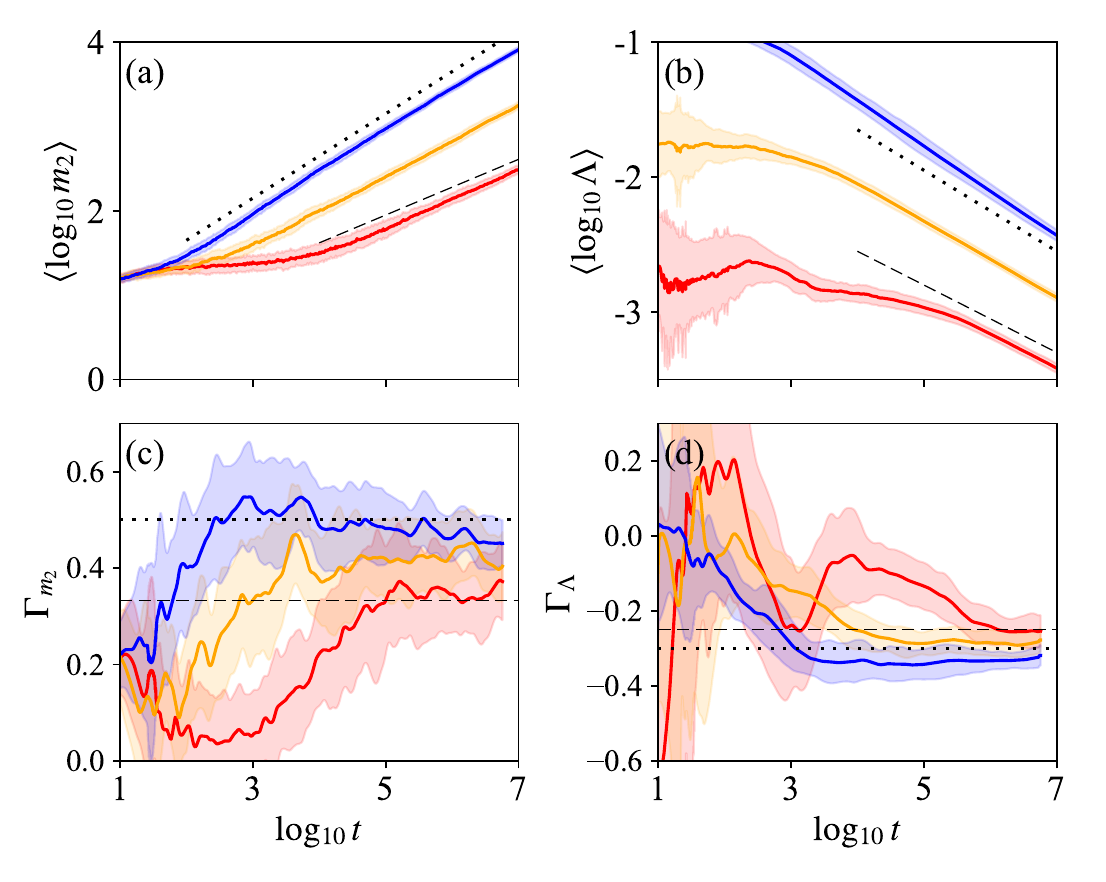}
    \caption{Averaged (and smoothed) results over 50 disorder realizations of the time evolution of the wave packet's (a) second moment $m_2(t)$ \eqref{m2},  and (b) ftMLE $\Lambda(t)$ \eqref{mle}, along with the related numerically computed [Eq.~\eqref{eq:rate}] derivatives (c) $\Gamma_{m_2}(t)$, and (d) $\Gamma_{\Lambda}(t)$, for cases $W_{v,1}$ (red curves), $W_{v,2}$ (orange curves) and $S_{v,3}$ (blue curves). The dashed (dotted) lines correspond to $\Gamma_{m_2}=0.33$ ($\Gamma_{m_2}=0.5$) in (a) and (c), and to $\Gamma_{\Lambda} =  -0.25$ ($\Gamma_{\Lambda} =  -0.3$) in (b) and (d).  In each panel the shaded area denotes one standard deviation error.}
    \label{fig:verti}
\end{figure}

We start our analysis by fixing $W=2.46$ and considering cases with increasing values of $\beta$, i.e., we move vertically in the parameter space of Fig.~\ref{fig:stub_param}. In particular, we consider the following three cases
\begin{enumerate}
  \item $W_{v,1}$:  $\beta = 0.02$, $W=2.46$, $L = 12$, $N=1001$, 
  \item $W_{v,2}$:  $\beta = 0.05$, $W=2.46$, $L = 12$, $N=1001$, 
  \item $S_{v,3}$: $\beta = 0.4$, $W=2.46$, $L = 12$, $N=1001$.
\end{enumerate}
We note that case $W_{v,1}$ (studied in Sect.~\ref{sec:weak_rep_cases}) is located well inside that weak chaos regime in Fig.~\ref{fig:stub_param}, while case $S_{v,3}$ (discussed in Sect.~\ref{sec:strong_rep_cases}) clearly belongs to the strong chaos regime. In Fig.~\ref{fig:verti},   we present the time evolution of  $\langle  \log_{10} m_2 \rangle$ [Fig.~\ref{fig:verti}(a)] and $\langle \log_{10}  \Lambda \rangle $ [Fig.~\ref{fig:verti}(b)], along with  their numerically computed derivatives $\Gamma_{m_2} (t)$ [Fig.~\ref{fig:verti}(c)] and $\Gamma_{\Lambda} (t)$ [Fig.~\ref{fig:verti}(d)] for cases $W_{v,1}$ (red curves), $W_{v,2}$ (orange curves) and $S_{v,3}$ (blue curves). The weak chaos dynamics, characterized by  $\Gamma_{m_2}=0.33$ [dashed lines in Figs.~\ref{fig:verti}(a) and \ref{fig:verti}(c)] is observed for the $W_{v,1}$ case, while the strong chaos behavior, corresponding to $\Gamma_{m_2}=0.5$ [dotted lines in Figs.~\ref{fig:verti}(a) and \ref{fig:verti}(c)] is clearly seen for the $S_{v,3}$ case. Increasing the nonlinear parameter from $\beta= 0.02$  ($W_{v,1}$ case) to $\beta= 0.05$  ($W_{v,2}$  case) moves the system to the transition region between the weak and strong chaos behaviors in Fig.~\ref{fig:stub_param}, and  results in  a faster spreading, which also starts earlier in time.  In particular,  the $\Gamma_{m_2}(t)$ exponent [orange curve in Fig.~\ref{fig:verti}(c)] reaches higher values than $\Gamma_{m_2}=0.33$ (weak chaos case $W_{v,1}$), which nevertheless remain smaller than $\Gamma_{m_2}=0.5$ (strong chaos case $W_{v,3}$). As we see in Fig.~\ref{fig:verti}(c) the $\Gamma_{m_2} (t)$ of the $W_{v,2}$ case eventually approaches values close to $\Gamma_{m_2}\approx 0.4$, a value which was also reported in Refs.~\cite{SHEPELYANSKYD.L1993Doqc,PikovskyAS2008DoAl}. The further increase of the nonlinearity strength to  $\beta=0.4$ (strong chaos case $S_{v,3}$) leads to an even faster spreading, which also sets in earlier in time, as the results of  Fig.~\ref{fig:verti}(a) clearly show. From Fig.~\ref{fig:verti}(b) we see that $\langle \log_{10}  \Lambda \rangle (t)$ decreases to zero following power law decays in all cases. The asymptotic values of the related exponents $\Gamma_{\Lambda} (t)$ change as we move from the weak chaos case $W_{v,1}$ [$\Gamma_{\Lambda} \approx -0.25$, dashed lines in Figs.~\ref{fig:verti}(b) and \ref{fig:verti}(d)], to the strong chaos case $S_{v,3}$ [$\Gamma_{\Lambda} \approx -0.3$, dotted lines in Figs.~\ref{fig:verti}(b) and \ref{fig:verti}(d)], with $-0.3 \lesssim \Gamma_{\Lambda} \lesssim -0.25 $ for the $W_{v,2}$ case [orange curve in Fig.~\ref{fig:verti}(d)] located near the transition region between the weak and strong chaos regimes in Fig.~\ref{fig:stub_param}.

\begin{figure}[t]
    \centering
    \includegraphics[scale=0.5]{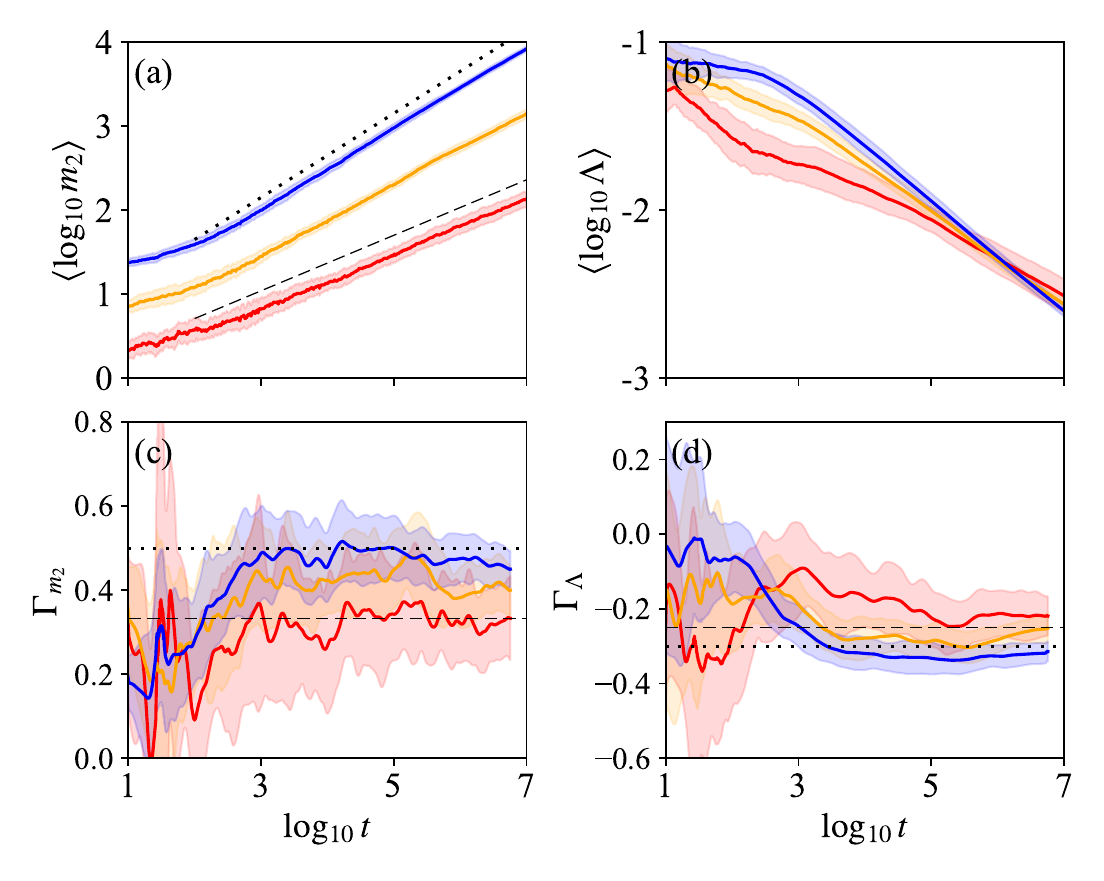}
    \caption{Similar to Fig.~\ref{fig:verti} but for cases $W_{h,1}$ (red curves), $W_{h,2}$ (orange curves), and $S_{h,3}$ (blue curves). The dashed (dotted) lines correspond to $\Gamma_{m_2}=0.33$ ($\Gamma_{m_2}=0.5$) in (a) and (c), and to $\Gamma_{\Lambda} =  -0.25$ ($\Gamma_{\Lambda} =  -0.3$) in (d).  }
    \label{fig:hori}
\end{figure}

Moving now horizontally in the parameter space $(W, \delta)$ of  Fig.~\ref{fig:stub_param},  we keep $\beta=0.25$ and decrease the value of $W$ (moving from larger to smaller disorder strengths) by considering the cases
\begin{enumerate}
  \item $W_{h,1}$:  $\beta = 0.25$, $W=5.2$, $L = 3$, $N=1001$, 
  \item $W_{h,2}$:  $\beta = 0.25$, $W=3.5$, $L = 6$, $N=1001$, 
  \item $S_{h,3}$:  $\beta = 0.25$, $W=2.2$, $L = 15$, $N=1001$.
\end{enumerate}
In Fig.~\ref{fig:hori},   we show the time evolution of  $\langle  \log_{10} m_2 \rangle (t)$ [Fig.~\ref{fig:hori}(a)] and its derivative $\Gamma_{m_2} (t)$ [Fig.~\ref{fig:hori}(c)], along with  $\langle \log_{10}  \Lambda \rangle (t)$ [Fig.~\ref{fig:hori}(b)] and its  derivative $\Gamma_{\Lambda} (t)$ [Fig.~\ref{fig:hori}(d)] for the $W_{h,1}$ (red curves), $W_{h,2}$ (orange curves) and $S_{h,3}$ (blue curves) cases. From the results of Fig.~\ref{fig:hori},  we see that the $W_{h,1}$ case, which is located well inside the weak chaos regime in the parameter space of Fig.~\ref{fig:stub_param}, exhibits weak chaos dynamics characterized by $\Gamma_{m_2}=0.3$ [dashed lines in Figs.~\ref{fig:hori}(a) and \ref{fig:hori}(c)]. On the other hand,  the $S_{h,3}$ case clearly shows the characteristics of strong chaos dynamics, with $\Gamma_{m_2} (t)$ reaching the value $\Gamma_{m_2}=0.5$ [dotted lines in  Figs.~\ref{fig:hori}(a) and \ref{fig:hori}(c)], followed by a slight decrease of its values for $t \gtrsim 10^5$  [Fig.~\ref{fig:hori}(c)]. Decreasing the value of the disorder strength from $W=5.2$ (weak chaos case  $W_{h,1}$) to  $W=3.5$ (case $W_{h,2}$), and eventually to  $W=2.2$ (strong chaos case $S_{h,3}$), we observe a gradual change in the values of $\Gamma_{m_2}$   with $\Gamma_{m_2} \approx 0.4 $ for the $W_{h,2}$ case [orange curve in Fig.~\ref{fig:hori}(c)], but we do not observe any significant change at the starting time of the wave packet spreading,  as all curves in Fig.~\ref{fig:hori}(a) start exhibiting a power law increase around the same time ($t \approx 10^2$).  In all considered cases,  the ftMLE [Eq.~\eqref{mle}]  exhibits a power law decay whose exponent $\Gamma_{\Lambda}$  varies from $\Gamma_{\Lambda} = -0.25$ [dashed line in Fig.~\ref{fig:hori}(d)],  for the weak  chaos case $W_{h,1}$ to $\Gamma_{\Lambda} = -0.3$ [dotted line in Fig.~\ref{fig:hori}(d)] for the strong chaos  case $S_{h,3}$, with the intermediate case $W_{h,2}$ located close to the transition zone between the weak and strong chaos regimes in Fig.~\ref{fig:stub_param},  characterized by $-0.3 \lesssim \Gamma_{\Lambda} \lesssim -0.25$.

In all  cases presented in Figs.~\ref{fig:verti} and \ref{fig:hori},  we observe subdiffusive wave packet spreading characterized by a power law increase of the second moment [Eq.~\eqref{m2}], i.e.~$m_2 \propto t^{\Gamma_{m_{2}}}$. In addition, in all these cases we found a power law decrease of the ftMLE [Eq.~\eqref{mle}], i.e.~$\Lambda \propto t^{\Gamma_{\Lambda}}$, with $\Gamma_{\Lambda} \neq -1$,  which is a value indicating  regular behavior.  For fixed values of $W$ the  increase of the nonlinear parameter $\beta$ leads to faster spreading and the transition from the weak chaos regime (characterized by $\Gamma_{m_{2}}=0.33$ and $\Gamma_{\Lambda}=-0.25$) to the strong chaos regime (having $\Gamma_{m_{2}}=0.5$ and $\Gamma_{\Lambda}=-0.3$),  something which is also observed when $W$ is decreased while $\beta$ is kept constant.  The transition between these two spreading regimes is  not abrupt, but happens in the rather gradual mode, which is reflected to the smooth change of the asymptotic $\Gamma_{m_{2}}$ and $\Gamma_{\Lambda}$  values, similarly to what was observed for the DKG and DDNLS systems \cite{BLSKF_11}.

\subsection{The effect of small frequency gaps}
\label{sec:freq_gap}

The weak chaos, strong chaos and self-trapping  cases of the stub lattice Hamiltonian [Eq.~\eqref{H_sl_normal}] we have considered so far were located at  the right side of the $W = 1.58$ vertical dotted line in the system's parameter space $(W, \delta)$  in  Fig.~\ref{fig:stub_param}. According to our discussion in Sect.~\ref{sec:model}, all these cases correspond to system arrangements for which no gap in the normal mode frequency spectrum is present.  

Although studying the influence of the frequency gaps on the dynamical behavior of the weak and strong chaos regimes is particularly interesting, this task is computationally demanding. In practice, the smallest disorder strength that allows wave packet propagation over sufficiently long times ($t = 10^{7}$) without boundary effects, within feasible CPU times, is $W = 1.16$. As a result, it is difficult to systematically explore the full range of gap widths ($0 \leq \alpha \leq 1$) that appear in the linear spectrum for $0 \leq W \lesssim 1.58$ (Figs.~\ref{fig:frequencyspectrum} and \ref{fig:disorderevolution}). Therefore, in the following, we restrict our analysis to the regime of small gaps, ($\alpha$ is small) focusing on the dynamical behavior close to the critical point where the gaps begin to close.

In order to investigate the potential influence of the appearance of these small gaps on the dynamical behavior of the weak and strong chaos spreading regimes, we consider for each dynamical regime four different parameter cases. In particular, we numerically  study  the wave packet propagation for the following four weak  chaos cases
\begin{enumerate}
  \item $W_{e,1}$:  $\beta = 0.03$, $W=1.94$, $L = 18$, $N=1001$, $\alpha = 0$, 
  \item $W_{e,2}$:  $\beta = 0.02$, $W=1.68$, $L = 24$, $N=1001$, $\alpha = 0$, 
  \item $W_{e,3}$:  $\beta = 0.01$, $W=1.42$, $L = 35$, $N=1501$, $\alpha = 0.007$, 
  \item $W_{e,4}$:  $\beta = 0.005$, $W=1.16$, $L = 53$ $N=1501$, $\alpha = 0.124$,
\end{enumerate}
as well as, for four parameter setups belonging to the strong chaos dynamical regime, namely
\begin{enumerate}
  \item $S_{e,1}$:  $\beta = 0.55$, $W=1.94$, $L = 18$, $N=1501$, $\alpha = 0$, 
  \item $S_{e,2}$:  $\beta = 0.5$, $W=1.68$, $L = 24$, $N=1501$, $\alpha = 0$, 
  \item $S_{e,3}$:  $\beta = 0.45$, $W=1.42$, $L = 35$, $N=1501$, $\alpha = 0.007$, 
  \item $S_{e,4}$:  $\beta = 0.4$, $W=1.16$, $L = 53$, $N=1801$, $\alpha = 0.124$.  
\end{enumerate}
We note that the above cases are ordered in decreasing value of their disorder strength $W$, namely for $W=1.94$ (cases $W_{e,1}$ and $S_{e,1}$), and $W=1.68$ (cases $W_{e,2}$ and $S_{e,2}$) for which no frequency gap  ($\alpha = 0$) is present, followed by $W=1.42$ (cases $W_{e,3}$ and $S_{e,3}$), and $W=1.16$ (cases $W_{e,4}$ and $S_{e,4}$), for which $0 < \alpha < 1$. From the results of Fig.~\ref{fig:volumepart}, we see that the localization volume  of the normal modes, or in other words their lattice extent,  is increased as $W$ decreases. Since in our numerical simulations we initially excite $L$ central sites, with $L$ being practically equal to  $\langle V \rangle$, $L$ increases as $W$ becomes smaller. Consequently,  in order to avoid the created wave packets reaching the lattice boundaries during our numerical simulation (something which would introduce boundary effects in the overall dynamics), we had to significantly increase the size of the considered lattices, reaching a number of $N=1801$ cells for the strong chaos case $S_{e,4}$. As mentioned above, the related value, $W = 1.16$, represents the smallest value of the disorder strength we could reach in our numerical simulations to achieve the propagation of wave packets for large enough times ($t=10^7$) without having any boundary effects introduced in the dynamics, in accessible CPU times. 

\begin{figure}[t]
    \centering
    \includegraphics[scale=0.5]{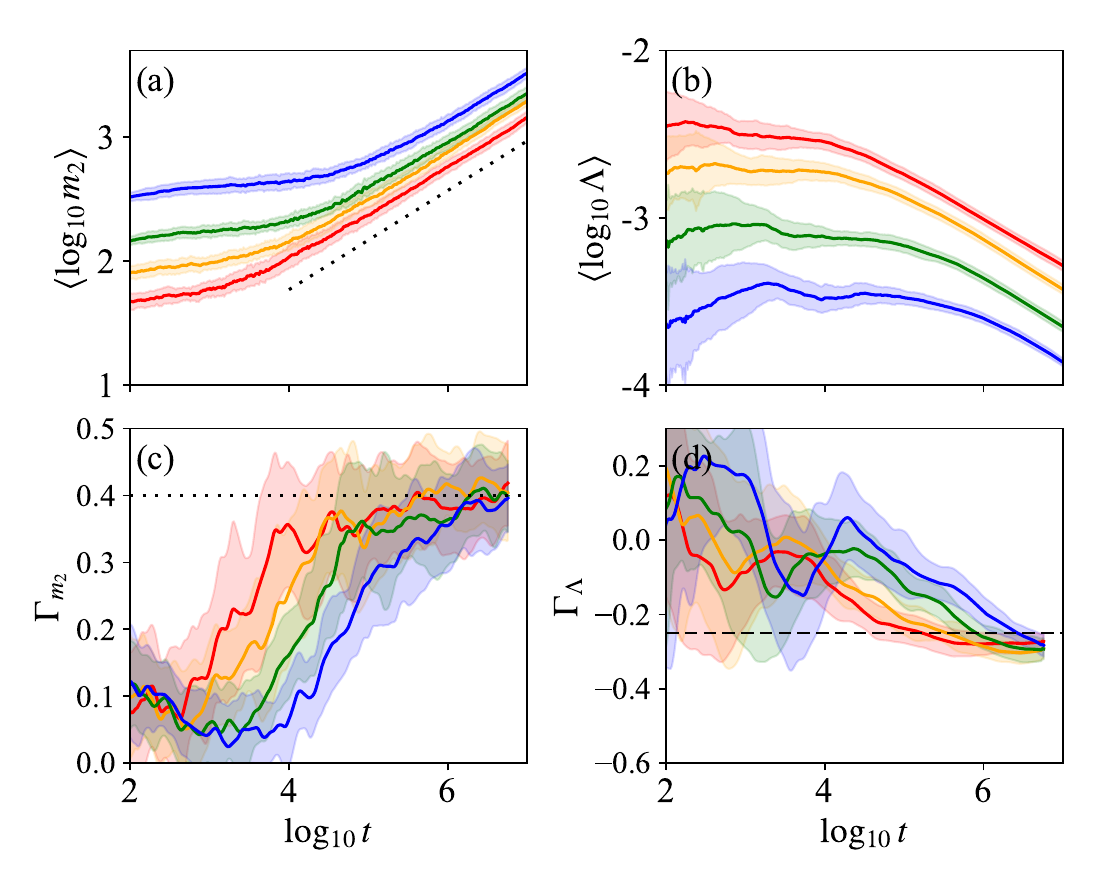}
    \caption{Similar to Fig.~\ref{fig:verti} but for the weak chaos cases $W_{e,1}$ (red curves), $W_{e,2}$ (orange curves), $W_{e,3}$ (green curves) and $W_{e,4}$ (blue curves). The  dotted lines  in (a) and (c) correspond to $\Gamma_{m_2}=0.4$, while the dashed line in (d) indicates $\Gamma_{\Lambda} =  -0.25$ }
    \label{fig:overallweak}
\end{figure}

In Fig.~\ref{fig:overallweak},  we present the time evolution of the averaged (over 50 realization) $m_2(t)$ [Fig.~\ref{fig:overallweak}(a)] and its derivative $\Gamma_{m_2}(t)$ [Fig.~\ref{fig:overallweak}(c)], as well as the average ftMLE $\Lambda(t)$  [Fig.~\ref{fig:overallweak}(b)] and its  derivative  $\Gamma_{\Lambda} (t)$ [Fig.~\ref{fig:overallweak}(d)] for the weak  chaos cases $W_{e,1}$ (red curves), $W_{e,2}$ (orange curves), $W_{e,3}$ (green curves) and $W_{e,4}$ (blue curves). By checking the location of these cases in the parameter space of  Fig.~\ref{fig:stub_param}, we see that they are not very far away from  the borderline separating the weak and strong chaos regimes. Thus, it is not surprising that their dynamical behavior is similar to what was observed for the $W_{v,2}$ case (orange curves in Fig.~\ref{fig:verti}) and  the $W_{h,2}$ case (orange curves in Fig.~\ref{fig:hori}). In particular, all cases exhibit a power law increase of the second moment ($m_2 \propto t^{\Gamma_{m_2}}$) with an exponent which eventually  becomes $\Gamma_{m_2} \approx 0.4$  [dotted lines in Figs.~\ref{fig:overallweak}(a) and \ref{fig:overallweak}(c)]. The time at which the power law increase of the second moment becomes apparent is growing as we move towards smaller values of $W$, i.e.~moving from the red curves to the blue curves in Figs.~\ref{fig:overallweak}(a) and \ref{fig:overallweak}(c). Furthermore, in all cases,  the average ftMLE $\Lambda$  decreases as $\Lambda \propto t^{\Gamma_{\Lambda}}$, with $\Gamma_{\Lambda}$ attaining a value  slightly lower than $\Gamma_{\Lambda}=-0.25$  [denoted by the horizontal straight dashed line in Figs.~\ref{fig:overallweak}(d)], similar to what was observed for the $W_{v,2}$  and $W_{h,2}$ cases  in respectively Figs.~\ref{fig:verti}(d) and  \ref{fig:hori}(d).  Thus, all four considered weak chaos cases show similar dynamical behaviors, both for the evolution of $m_2$ and $\Lambda$, exhibiting practically the same exponents in their power law time evolutions. We note that the asymptotic values of these exponents are reached at  later times for smaller $W$ values. Due to the smooth variation of the curves shown in Fig.~\ref{fig:overallweak}, as we move from the $W_{e,4}$ case (with a frequency gap of $\alpha = 0.124$) through the critical point where the gaps begin to vanish, and finally to the $W_{e,1}$ case where no gap exists, we do not observe any significant effect of the gap closure on the dynamics within the weak chaos spreading regime.

\begin{figure}[t]
    \centering
    \includegraphics[scale=0.5]{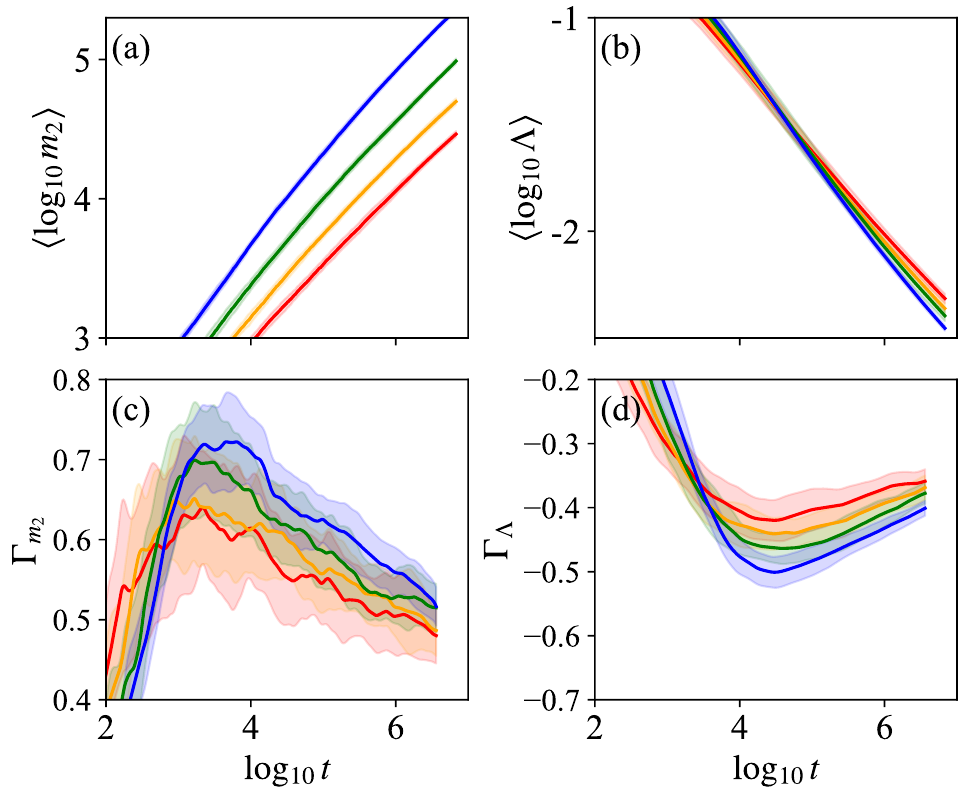}
    \caption{Similar to Fig.~\ref{fig:verti} but for the strong chaos cases $S_{e,1}$ (red curves), $S_{e,2}$  (orange curves), $S_{e,3}$ (green curves) and $S_{e,4}$ (blue curves). }
    \label{fig:overallstrong}
\end{figure}

Analogous results to the ones presented in Fig.~\ref{fig:overallweak} are shown  in Fig.~\ref{fig:overallstrong} for  the strong chaos cases  $S_{e,1}$ (red curves), $S_{e,2}$ (orange curves), $S_{e,3}$ (green curves) and $S_{e,4}$ (blue curves). We note  that, although all presented curves have similar shapes,  no clear asymptotic dynamical behavior can be obtained. The average $m_2(t)$ of all considered cases grows in time, achieving larger values for smaller $W$ [Fig.~\ref{fig:overallstrong}(a)].  The corresponding derivatives $\Gamma_{m_2}(t)$ [Fig.~\ref{fig:overallstrong}(c)]  show a sharp increase for small times reaching their highest values in the interval $10^3 \lesssim t \lesssim 10^4$, showing a clear tendency to decrease to values close to $\Gamma_{m_2}=0.5$, i.e.~the value observed for  the strong chaos cases  $S_{v,3}$ [Fig.~\ref{fig:verti}(c)] and  $S_{h,3}$ [Fig.~\ref{fig:hori}(c)] , although the potential  saturation to this value is not clear up to the numerically accessible integration time $t=10^7$.  In a similar manner, $\Lambda(t)$ shows a clear decrease in time  for all cases [Fig.~\ref{fig:overallstrong}(b)], although the related derivatives  $\Gamma_{\Lambda}(t)$ did not manage to saturate to well defined values  [Fig.~\ref{fig:overallstrong}(d)] until the considered integration time. As there is no clear difference in Fig.~\ref{fig:overallstrong}  between the  results obtained for the strong chaos cases having no frequency gap (cases $S_{e,1}$ and $S_{e,2}$), and the ones where the small gap is present (cases $S_{e,3}$ and $S_{e,4}$),  we again do not see signs of any impact of small frequency gaps close to the critical $W = 1.58$ value on the dynamics of the strong chaos regime. Nevertheless, we have to stress that, most probably, longer simulations are needed in order for the features of the strong chaos cases (e.g.~the $\Gamma_{m_2}$ and $\Gamma_{\Lambda}$ values) to settle to some well define behaviors. 

We emphasize again that the results presented in this section for both the weak and strong chaos cases, due to computational constraints, correspond to conditions near the critical disorder value at which frequency gaps start to disappear and $\alpha$ is small. Greater computational resources would be required to investigate the behavior of the different regimes in the presence of a larger frequency gaps.

\section{Summary and conclusions}
\label{sec:conclusion}

In this work we numerically investigated the interplay between nonlinearity, disorder and the presence of FBs and gaps in the frequency spectrum of a tight-binding lattice model, namely the stub lattice of Eq.~\eqref{H_sl_normal}, which consists of coupled cells each one of which contained three interconnected sites. In particular, we performed extensive numerical simulations of the evolution of initially localized  excitations at the middle of this lattice, for various parameter sets of the system, and studied the dynamical properties of the propagation of the created wave packets, i.e.~norm distributions. 

The analysis of the linear stub lattice  showed the existence of a FB for the ordered version of the system, which was separated by well defined gaps from two dispersive frequency bands. The introduction of disorder distorted the  FB and reduced the size of these  gaps. Actually, the frequency gaps disappeared when the disorder strength $W$ became $W \gtrsim 1.58$. Furthermore, the inclusion of nonlinearity led to the interaction of the system's normal modes and the appearance of chaotic behavior.

In our numerical simulations we followed the propagation of wave packets by implementing a symplectic integration  scheme, and quantified the wave packet's extent by computing its second moment $m_2$ [Eq.~\eqref{m2}],  and its degree of localization by estimating the number of highly excited cells through the evaluation of the participation number $P$ [Eq.~\eqref{P}]. In addition, we quantified the strength of chaos by measuring the ftMLE $\Lambda$ [Eq.~\eqref{mle}].

We theoretically predicted, and numerically verified the existence of three different dynamical regimes exhibiting distinct features, namely the weak and strong chaos spreading regimes, for which subdiffusive spreading of the created wave packet was observed, and the so-called self-trapping regime, where the largest part of the wave packet remained localized near the region of the initial excitation. In addition, we identified the location of these dynamical regimes in the system's parameter space (Fig.~\ref{fig:stub_param}) and investigated in detail their characteristics. We observed that the time involution of $m_2$, $P$ and $\Lambda$ were well approximated by power laws ($m_2 \propto t^{\Gamma_{m_2}}$, $P \propto t^{\Gamma_{P}}$, $\Lambda \propto t^{\Gamma_{\Lambda}}$) when  wave packet spreading was observed. In particular, we found  that the  exponents of these power laws asymptotically tended to $\Gamma_{m_2}=0.33$, $\Gamma_{P}=0.167$ and $\Gamma_{\Lambda}=-0.25$  for the weak chaos  case, and to $\Gamma_{m_2}=0.5$, $\Gamma_{P}=0.25$ and $\Gamma_{\Lambda}=-0.3$ for the strong chaos spreading regime. It is worth noting that the three dynamical regime observed in our work, along with the specific values of the various power law exponents  we found, were also reported in several studies of the DKG and DDNLS models \cite{FlachS2009Usow,SkokosCh2009Dowp,LaptyevaT.V2010Tcfs,BLSKF_11,SkokosCh2013Ncod,SenyangeB2018Coce}. Furthermore, similarly to what has been reported for the DKG and DDNLS models, we showed that the boundary between the weak and the strong chaos spreading regimes in the system's parameter space was not  abrupt, but the transition between these regimes happened in the rather gradual manner, characterized by smooth changes in the asymptotic values of the various power law exponents. 

Another  feature that the stub lattice model has in common with the DKG and DDNLS systems, is that the number of highly chaotic cells (the so-called chaotic hot spots), whose position is identified as the part of the lattice where the DVD has large values, remain rather small throughout the evolution of the wave bucket, something which is reflected on the rather localized and pointy shape of the DVD.  Furthermore, the DVD, which remains always inside the excited part of the lattice, exhibits oscillations whose amplitudes increase in time. This meandering of the chaotic hot spots supports the homogenization of the wave packet's chaoticity.

The detailed study of various cases located at different regions of the system's parameter space, showed that the increase of the nonlinearity strength led to faster spreading, which in addition started earlier in time, while spreading was delayed when the disorder strength was increased. 
Another important outcome of our work is that we did not find any clear evidence of a significant effect of the frequency gaps on the dynamical features of the weak and strong chaos regimes. In practice, our analysis was limited to small gap values due to computational constraints, as discussed in Sect.~\ref{sec:freq_gap}. The weak and strong chaos cases we examined, with disorder strengths $W$ both above (no frequency gaps present) and below (small gaps appearing in the spectrum) the threshold value $W = 1.58$, exhibit similar dynamical behaviors. However, in the strong chaos cases, well-defined asymptotic behaviors were not fully reached within the considered integration times.

Our results provide some valuable insights into the energy propagation in FB lattices in the presence of disorder and nonlinearity. The fact that similar mechanisms and behaviors to the ones found for the DKG and DDNLS models, were also observed in the stub lattice system studied here (namely, the existence of the weak and strong chaos spreading regimes, the characteristics of the wave packet propagation in these regimes which is described by   well defined power law  evolutions of the $m_2$, $P$, and $\Lambda$ in time, the enhancement of the wave packet's chaoticity  through the meandering of chaotic hot spots inside it,  along with the appearance of the self-trapping  regime), suggest the generality of the reported behaviors  for a variety of disordered systems,  including FB lattices. Nevertheless,  additional investigations of the chaoticity and the properties of wave packet propagation in different disordered, nonlinear FB models are needed in order to further consolidate the universality of these behaviors. This is a task we plan to undertake in the future.

\section*{Acknowledgements}

A.~N.~acknowledges support from the Research Committee of the University of Cape Town (UCT). Ch.~S.~thanks the Le Mans Universit\'e  for its hospitality during his visits, when parts of this work were carried out, and acknowledges support by the UCT’s Research Committee (URC). We thank the High Performance Computing facility of UCT and the Centre for High Performance Computing \cite{chpc} of South Africa for providing computational resources for this project.


\section*{Author Declarations}
The authors have no conflicts to disclose.

\section*{Data Availability Statement}

The data that support the findings of this study are available within the article, and upon reasonable request.

\appendix
\section{Numerical setup and approach}
\label{APPENDIX:A1}

The numerical approach taken in Sect.~\ref{sec:Numerical Results} follows the one described in the appendix of Ref.~\cite{BLSKF_11}. In particular,  we perform two successive two part splits of Hamiltonian in Eq.~\eqref{H_sl_normal}. We write $H = \mathcal{A} + \mathcal{B}$, with $\mathcal{A}$, comprising the first six terms of Eq.~\eqref{H_sl_normal}, being integrable, while $\mathcal{B}$ contains the last three terms of Eq.~\eqref{H_sl_normal}. Since $\mathcal{B}$ is not integrable we further split it into two integrable parts so that  $\mathcal{B} \left( q_{n}^{(K)}, p_{n}^{(K)}\right) = \mathcal{P}\left( p_{n}^{(K)}\right) + \mathcal{Q}\left( q_{n}^{(K)}\right)$. In our study we use fixed boundary conditions $q_0^{(K)} = q_{N+1}^{(K)} = p_0^{(K)} = p_{N+1}^{(K)} = 0$ (with $K$ denoting A, B and C sites), performing, in most cases, numerical simulations up to a final integration time of $t_f = 10^7$. In order to avoid finite-size  effects we ensure that in all examined cases  the wave packet does not reach the boundaries of the studied lattice, by considering large enough lattice sizes $N\approx 500 - 2200$. The used integration time steps, $\tau \approx 0.1 - 0.2$, led to a very good conservation of the system's integrals of motion, as the absolute value of the energy relative error was typically kept smaller than $10^{-5}$, and that of the  norm relative error was  below $10^{-4}$.

\nocite{*}

\bibliography{apssamp}

\end{document}